\documentclass[12pt,letterpaper]{article}
\pdfoutput=1
\usepackage[colorlinks=true,
linkcolor=Red, 
citecolor=Red,
filecolor=Red,
urlcolor=Red,
linktoc=page, 
pdfstartview=FitV,
bookmarksopen=true]{hyperref}
\usepackage[left=2cm,top=1cm,right=3cm,nohead]{geometry}
\usepackage[dvipsnames]{xcolor}
\usepackage{colortbl}
\usepackage[utf8]{inputenc} 
\usepackage{color}
\usepackage{epsfig,latexsym,amsfonts,mathtools,amsthm,amssymb,amsbsy,multirow,slashed,
	mathrsfs,wasysym,textcomp,subfigure,wrapfig,comment,bbold,array,longtable,multirow,setspace,amsmath,pifont, enumitem}
\usepackage{cleveref}
\usepackage{tabularx}
\usepackage{graphicx}
\usepackage{setspace}
\usepackage{subfigure}
\usepackage{multirow}
\usepackage[bf]{caption}
\usepackage{braket}
\usepackage{comment}
\usepackage{float}
\usepackage{tikz}
\usetikzlibrary{mindmap,trees,shadows}
\colorlet{color1}{gray!25}
\usetikzlibrary{positioning}
\usetikzlibrary{intersections} 

\newlength{\PicScale}
\usepackage[UKenglish]{babel}
\usepackage[toc,page]{appendix}
\usepackage{hhline}
\usepackage{geometry}
\usepackage{cite}
\usepackage{datetime}
\usepackage{bbm} 
\usepackage{bm} 
\hypersetup{
	pdftitle={},
	pdfauthor={},
	pdfsubject={}
} 
\usepackage{subfiles}
\definecolor{Gray}{gray}{0.94}

\numberwithin{equation}{section}

\makeatletter
\def\@cline#1-#2\@nil{
	\omit
	\@multicnt#1
	\advance\@multispan\m@ne
	\ifnum\@multicnt=\@ne\@firstofone{&\omit}\fi
	\@multicnt#2
	\advance\@multicnt-#1
	\advance\@multispan\@ne
	\leaders\hrule\@height\arrayrulewidth\hfill
	\cr
	\noalign{\nobreak\vskip-\arrayrulewidth}}
\makeatother

\graphicspath{{./Images/}}

\begin{document}
\pagestyle{empty}
\begin{center}        
  {\bf\LARGE Unifying the 6D $\mathcal{N}=(1,1)$ String Landscape  \\ [3mm]}

\setstretch{1.3}
\large{ Bernardo Fraiman$^{*\,\dagger}$ and H\'ector Parra De Freitas$^{*}$
	\\[2mm]}
{\small $*$  Institut de Physique Th\'eorique, Universit\'e Paris Saclay, CEA, CNRS\\ [-2mm]}
{\small\it  Orme des Merisiers, 91191 Gif-sur-Yvette CEDEX, France.\\[0.2cm] } 
{\small  $\dagger$ Instituto de Astronom\'ia y F\'isica del Espacio (IAFE-CONICET-UBA)\\ [-2mm]}
{\small \hspace{1em}  Departamento de F\'isica, FCEyN, Universidad de Buenos Aires (UBA) \\ [-2mm]}
{\small\it Ciudad Universitaria, Pabell\'on 1, 1428 Buenos Aires, Argentina\\ }

{\small \verb"bfraiman@iafe.uba.ar, hector.parradefreitas@ipht.fr"\\[-3mm]}
\vspace{0.3in}

\small{\bf Abstract} \\[3mm]\end{center}
We propose an organizing principle for string theory moduli spaces in six dimensions with $\mathcal{N} = (1,1)$, based on a rank reduction map, into which all known constructions fit. In the case of cyclic orbifolds, which are the main focus of the paper, we make an explicit connection with meromorphic 2D (s)CFTs with $c = 24$ ($c = 12$) and show how these encode every possible gauge symmetry enhancement in their associated 6D theories.  These results generalize naturally to non-cyclic orbifolds, into which the only known string construction (to our awareness) also fits. This framework suggests the existence of a total of 47 moduli spaces: the Narain moduli space, 23 of cyclic orbifold type and 23 of non-cyclic type. Of these only 17 have known string constructions. Among the 30 new moduli spaces, 15 correspond to pure supergravity, for a total of 16 such spaces. A full classification of nonabelian gauge symmetries is given, and as a byproduct we complete the one for seven dimensions, in which only those of theories with heterotic descriptions were known exhaustively.  

\newpage



\setcounter{page}{1}
\pagestyle{plain}
\renewcommand{\thefootnote}{\arabic{footnote}}
\setcounter{footnote}{0}

\tableofcontents	
\newpage

\section{Introduction}\label{s:Intro}
There has been some recent progress in understanding string theory in the regime with sixteen supercharges, which serves to test various ideas that could be generalized to other regimes. Some of the most prominent amongst these \cite{Kim:2019ths,Montero:2020icj,Hamada:2021bbz,Bedroya:2021fbu,Cvetic:2020kuw,Lee:2021usk,Lee:2021qkx,Collazuol:2022jiy} come from the Swampland program \cite{Vafa:2005ui,Ooguri:2006in}, in which many conjectures about the consistency of quantum gravity theories are made; particularly relevant to this work is the idea that string theory is a universal description of theories of quantum gravity (the so-called String Lamppost Principle.) Many other applications related to this work can be found in \cite{Kachru:2016ttg,Gaberdiel:2011fg,Cheng:2016org,Harrison:2020wxl,Harrison:2021gnp,Persson:2015jka,Persson:2017lkn}, which roughly speaking center around the appearance of moonshine in string theory.

It has been noted for example that the full moduli space of string vacua with $\mathcal{N} = 1$ in dimensions 9, 8 and 7 was missing various connected components associated to string theories with discrete theta angles \cite{MMHP}. This was predicted by noting that M-Theory compactifications on K3 surfaces with possible frozen singularities encompassed all known moduli space components given by other stringy descriptions\cite{deBoer:2001wca}, and that there were three more possibilities not accounted for in the literature \cite{HP}, one of which naturally lifts to eight and nine dimensions.    

On the other hand there has been some interest in the patterns of gauge symmetry enhancement in these theories, from the point of view of heterotic strings in  \cite{Fraiman:2018ebo,Font:2020rsk,Font:2021uyw,Fraiman:2021hma,Fraiman:2021soq,Cvetic:2021sjm} and Type IIB string junctions in \cite{Cvetic:2022uuu}. It has been shown in particular that from the gauge symmetry groups appearing in the Narain component (e.g.\ heterotic strings on tori), one can obtain those in the other moduli space components by a suitable rank reduction map.

The goal of this paper is to determine at once what are all the possible components in the moduli space of string vacua in six dimensions with $\mathcal{N} = (1,1)$ and what are all the possible nonabelian gauge symmetries that arise in them. The key idea is that the map which relates the Narain component to the others is known to be the same that relates moduli spaces of flat $G$-bundles on $T^2$ with $G$ a compact non-simply-connected Lie group \cite{Schweigert:1996tg,Lerche:1997rr,Fuchs:1995zr} in the cases where we have explicit control over the computation of gauge groups\footnote{By this we mean that we are able to extract the information of the gauge group of a given string vacuum from an associated lattice embedded into the overall charge lattice of the theory.} \cite{Fraiman:2021hma}. Roughly speaking, for a non-simply-connected gauge group $G$ in the Narain component, each order $n$ element in the fundamental group $\pi_1(G)$ defines a mapping to another gauge group $G'$. For five types of mapping, where the orders are $2,3,4,5,6$, we know that $G'$ belongs respectively to a heterotic $\mathbb{Z}_n$-triple of \cite{deBoer:2001wca}. 

There are however other possible mappings, many of which, as we will show, correspond to other known components in the moduli space which are described by cyclic orbifolds. This is achieved with the help of a relation between these components and meromorphic (s)CFTs with $c = 24$ ($c = 12$) that appear in the worldsheet CFTs of their $T^4$ compactifications, i.e.\ in 2D.  In turn, the classification of such CFTs \cite{Schellekens:1992db,Harrison:2020wxl} allows to predict the existence of eight more moduli space components given by cyclic orbifolds. An explicit connection between the string theories and this classification is given along the lines of Höhn's reconstruction of the current algebras in Schellekens' list \cite{2017arXiv170805990H}, i.e.\ the case $c = 24$, although we show that this construction, with a minor amendment, also gives those for the $c = 12$ sCFTs. 

We also show that the Niemeier lattices with roots encode every possible gauge symmetry enhancement in the Narain component, and combining this with the map to other components we find that all the enhancements in them are encoded by the so-called orbit lattices. These results are checked against those of \cite{Fraiman:2021hma} with perfect agreement, but we emphasize here that we have access to many more components. 

More speculatively but also completely naturally, we propose that the rank reduction map may be applied iteratively until the resulting gauge group is simply-connected, and that this exhausts every possible nonabelian gauge symmetry group for a 6D string vacuum with $\mathcal{N} = (1,1)$. Encouragingly, the $\mathbb{Z}_2 \times \mathbb{Z}_2$ asymmetric orbifold of \cite{deBoer:2001wca} corresponds precisely to a case of applying the rank reduction two times, but there are still other 22 potential constructions not accounted for in the literature, which we believe to be realizable as non-cyclic orbifolds. Remarkably, four of these exhibit the novel property of having odd gauge group rank; two have rank 5, one has rank 3 and another rank 1. This extends the possibler rank reductions $r$ in 6D theories with 16 supercharges to 
\begin{equation}
    r = 8, 12, 14, 15, 16, 17, 18, 19, 20\,.
\end{equation}
In the end we are led to a picture in which the complete moduli space consists of 47 connected components, of which 30 have no known string theory description. These are recorded in Tables \ref{tab:chargecyclic} and \ref{tab:chargenoncyclic} of Appendix \ref{app:lattices}. The gauge symmetry enhancements for every component are recorded in an online database \cite{fp2022}.

We start in Section \ref{s:Narain} with a treatment of the Narain component which we generalize in Section \ref{s:CHL} to the CHL string. In Section \ref{s:General} we present the generalization to every component of cyclic orbifold type; the rest of the (non-cyclic) cases are considered in \ref{s:Overall}. Section \ref{s:Conclusions} presents our conclusions.

\section{Narain Component}\label{s:Narain}
In this section we recall the basic aspects of symmetry enhancement in the Narain component of the moduli space of six dimensional $\mathcal{N}=(1,1)$ string vacua in the language of lattices. We show how to obtain every possible enhancement using the Niemeier lattices, and interpret this result in terms of the structure of the theory when compactified further on $T^4$.

\subsection{Preliminaries}
The simplest string vacua in six dimensions with $\mathcal{N} = (1,1)$ supersymmetry are described by any of the heterotic strings compactified on $T^4$ or, equivalently, type IIA strings compactified on K3. They live in the Narain moduli space
\begin{equation}
    \mathcal{M} \simeq O(4,20,\mathbb{Z}) \backslash O(4,20,\mathbb{R}) / O(4,\mathbb{R})\times O(20, \mathbb{R})\,,
\end{equation}
where the dilaton contribution $\mathbb{R}^+$ is omitted. The discrete group $O(4,20,\mathbb{Z})$ is the T-duality group of the theory in the heterotic string description, and corresponds to the automorphism group of the Narain lattice $\Gamma_{4,20}$. The symmetric space which it quotients is the Grassmannian with signature $(4,20)$. In other words,
\begin{equation}
    \mathcal{M} \simeq O(\Gamma_{4,20}) \backslash \text{Gr}(4,20)\,.
\end{equation}
Points in this space may be interpreted as the possible orientations of a negative definite $4$-plane relative to the lattice $\Gamma_{4,20}$, both embedded into $\mathbb{R}^{4,20}$. 

At special points in $\mathcal{M}$, the orthogonal complement of the 4-plane intersects a positive definite sublattice $W$ of $\Gamma_{4,20}$, with rank $r = 1,...,20$. These sublattices are primitive, which means that the intersection of their real span $W \otimes \mathbb{R}$ with $\Gamma_{4,20}$ is $W$ itself. Of these, some enjoy the property of having a root sublattice (a lattice generated by vectors with norm $v\cdot v = 2$) of maximal rank, which means that they are \textit{Lie algebra lattices}, and can be interpreted as the weight lattices of certain nonabelian simply-laced Lie groups $G$. In fact, the physics of the theory works out such that at these points the gauge symmetry of the vacua have nonabelian part exactly given by $G$.\footnote{In this paper we ignore the contributions to the gauge group coming from the gravity multiplet, as well as discrete factors such as found in $E_8 \times E_8 \rtimes \mathbb{Z}_2$.} In general we have the result that, at a given point in moduli space,
\begin{equation}\label{latticegauge}
    W_{G} \hookrightarrow \Gamma_{4,20} ~~~~~ \Leftrightarrow ~~~~~\text{Gauge group} = G\times U(1)^{20-r}\,,
\end{equation}
where $W_G$ is the weight lattice of $G$ with rank $r$ and its embedding into $\Gamma_{4,20}$ is primitive. Generic points have abelian symmetry group, and so the appearance of nonabelian factors is referred to as symmetry enhancement. For $r = 20$, the enhancement is said to be maximal.

It may be helpful to elaborate on the meaning of $W_G$. We take it to be the weight lattice in the same sense as one usually refers to the lattice $D_{16}$ extended by the positive spinor class $(\tfrac12,...,\tfrac12)$ as the weight lattice of $Spin(32)/\mathbb{Z}_2$. For any primitive embedding $W_G \hookrightarrow \Gamma_{4,20}$, the self-duality of the host charge lattice ensures that its projection onto the real space spanned by $W_G$ is $W_G^*$ \cite{Narain:1986qm}. Since the spectrum of the theory is complete, $W_G^*$ encodes the allowed charges under the gauge symmetry group ensuring that the fundamental group $\pi_1(G)$ is exactly the quotient of $W_G$ by its root sublattice, as should be for a simply-laced group. This picture breaks down and must be generalized for theories with non-self-dual charge lattices. A treatment of this problem can be found in \cite{Cvetic:2021sjm}. 

Relation \eqref{latticegauge} allows to study symmetry enhancements from a purely lattice-theoretical point of view, for which many tools are available. This is specially true for the case at hand since $\Gamma_{4,20}$ is an even self-dual lattice. Even more, the fact that the rank of this lattice is 24, which is a particularity of six dimensional theories, makes it so that the problem of determining every possible $W_G$ is exactly solvable without too much effort. Let us show how this is done.

\subsection{Symmetry enhancements from Niemeier lattices}

Euclidean even self-dual lattices exist only when their rank is a multiple of 8. Of rank 24 there exist 24 such lattices, of which 23 are Lie algebra lattices while the other has no roots at all. The former are known as \textit{Niemeier lattices} and will be central to our analysis; they are briefly described in Appendix \ref{app:niemeier}. The later, known as the \textit{Leech lattice}, will not be relevant for us. Indeed, our interest is in nonabelian gauge symmetry enhancements, i.e.\ Lie algebra lattices. 

To get a sense of how Niemeier lattices enter into the discussion of symmetry enhancements, take the lattice $N_\gamma = 3\, E_8$ and draw its Dynkin diagram:
\begin{eqnarray}
\begin{aligned}
\begin{tikzpicture}[scale = 1.2]
\draw(0,0)--(2.5,0);
\draw(0.5,0)--(0.5,1);
\draw[fill=white](0,0) circle (0.1);
\draw[fill=white](0.5,0) circle (0.1);
\draw[fill=white](0.5,0.5) circle (0.1);
\draw[fill=white](0.5,1) circle (0.1);
\draw[fill=white](1,0) circle (0.1);
\draw[fill=white](1.5,0) circle (0.1);
\draw[fill=white](2,0) circle (0.1);
\draw[fill=white](2.5,0) circle (0.1);

\begin{scope}[shift={(3,0)}]
\draw(0,0)--(2.5,0);
\draw(0.5,0)--(0.5,1);
\draw[fill=white](0,0) circle (0.1);
\draw[fill=white](0.5,0) circle (0.1);
\draw[fill=white](0.5,0.5) circle (0.1);
\draw[fill=white](0.5,1) circle (0.1);
\draw[fill=white](1,0) circle (0.1);
\draw[fill=white](1.5,0) circle (0.1);
\draw[fill=white](2,0) circle (0.1);
\draw[fill=white](2.5,0) circle (0.1);
\end{scope}

\begin{scope}[shift={(6,0)}]
\draw(0,0)--(2.5,0);
\draw(0.5,0)--(0.5,1);
\draw[fill=white](0,0) circle (0.1);
\draw[fill=white](0.5,0) circle (0.1);
\draw[fill=white](0.5,0.5) circle (0.1);
\draw[fill=white](0.5,1) circle (0.1);
\draw[fill=white](1,0) circle (0.1);
\draw[fill=white](1.5,0) circle (0.1);
\draw[fill=white](2,0) circle (0.1);
\draw[fill=white](2.5,0) circle (0.1);
\end{scope}
\end{tikzpicture}
\end{aligned}
\end{eqnarray}
Deleting a node in this diagram selects a rank 23 primitive sublattice $W_{23}\hookrightarrow N_\gamma$, e.g.\ $W_{23} = 2\, E_8 \oplus E_7$. If one interprets $N_\gamma$ as the weight lattice of the gauge group $G_{N_\gamma} = E_8^3$, this procedure is equivalent to moving away from a point of symmetry enhancement in moduli space (as we will see, this scenario does appear in two spacetime dimensions where the gauge groups have rank 24). Repeating this procedure a total of four times, we are left with a rank 20 weight lattice for some gauge group (in this case a root lattice since the gauge group is simply connected). 

Our claim is that the procedure just outlined, if done in all possible ways for all Niemeier lattices, will produce a complete list of the maximal symmetry enhancements of the Narain component discussed in the previous section. This will be proven shortly. It is however instructive to note first that after deleting four nodes arbitrarily in one of the $E_8$ sublattices of $N_\gamma$, the result is a gauge group of the form 
\begin{equation*}
    G = E_8^2 \times G_{(4)}\,,
\end{equation*}
where $G_{(4)}$ is an arbitrary simply-laced compact Lie group of rank 4. It is well known that such a gauge group can be obtained in the $E_8\times E_8$ heterotic string on $T^4$ with null Wilson lines and appropriate values for the metric and B-field (see e.g.\ \cite{Ginsparg:1986bx}). The analogous case of the $Spin(32)/\mathbb{Z}_2$ heterotic string is to be found by deleting four nodes in the $E_8$ sublattice of $N_\beta = W_{Spin(32)/\mathbb{Z}_2} \oplus E_8$.

In the language of lattice embeddings, our claim is based on two statements:
\begin{enumerate}
    \item Any Lie algebra lattice $W_G$ of rank 20 which embeds primitively into $\Gamma_{4,20}$ also embeds primitively into some Niemeier lattice $N_I$. The converse is also true.
    
    \item The root sublattice of $W_G$ can be made to correspond to a rank 20 subdiagram of the Dynkin diagram of the Niemeier lattice into which it is embedded.
\end{enumerate}
The first of these two statements follows directly as a corollary of theorems 1 and 2 of \cite{Cheng:2016org} (recorded in Appendix \ref{app:NarNie},) by virtue of $W_G$ being a lattice whose embedding corresponds to a point in moduli space which is completely fixed by a subgroup of $O(\Gamma_{4,20})$ (the Weyl group of the associated root sublattice). The second statement follows from the possibility of matching the generating roots of the root sublattice of $W_G$ with a subset of those of $N_I$ by applying an appropriate transformation in the Weyl group of $N_I$. 

This result can be generalized to every other symmetry enhancement, not necessarily maximal, by observing that all possible gauge groups with rank $< 20$ are obtained from those of rank $20$ by deleting nodes. This is in contrast to compactifications to seven or more spacetime dimensions, where there is a special nonabelian gauge group of rank 16 which does not admit further enhancements \cite{Polchinski:1995df}. It is of the form
\begin{equation}
    G = ~\frac{Spin(16)^2}{\mathbb{Z}_2}\,, ~~~ \frac{Spin(8)^4}{\mathbb{Z}_2^2}\,, ~~~ \frac{Spin(4)^8}{\mathbb{Z}_2^5}\,, ~~~~~~~~~~\text{resp.}~d = 9,~8,~7\,.
\end{equation}
As is easy to infer, the next in the sequence is abelian.

It is satisfying that applying this method we obtain a list of gauge groups in perfect agreement with the results of \cite{Fraiman:2021hma}, which were obtained by means of an algorithm which finds embeddings of lattices $W_G$ into $\Gamma_{4,20}$ in an exploratory manner developed in \cite{Font:2020rsk,Font:2021uyw,Fraiman:2021soq}. This gives evidence for the effectiveness of the exploration algorithm, which has the added advantage of producing values for the moduli for which the enhancements occur and is not restricted to the six dimensional case.

\subsection{Connection to the 2D theory}\label{ss:Narain2D}
A physical interpretation is available for the results just reported in considering a further compactification of the theory on $T^4$ down to two dimensions. From the point of view of the heterotic string on $T^8$, the Narain lattice is now $\Gamma_{8,24}$. Its uniqueness as an even self-dual lattice with signature $(8,24)$ implies that any other such lattice is isomorphic to it. In particular, we have the isomorphisms
\begin{equation}\label{NarainIso}
    \Gamma_{8,24} \simeq N_I \oplus E_8(\text{-}1)\,.
\end{equation}
This means that there are 23 points (ignoring one associated to the Leech lattice) in the moduli space where the Narain lattice splits into two Euclidean lattices (here we take the active interpretation of the moduli space as that of polarizations of the Narain lattice, where the fixed negative definite plane corresponds to the subspace $\mathbb{R}^{8,0} \subset \mathbb{R}^{8,24}$). At these points, the gauge symmetry group is precisely the one whose weight lattice is $N_I$.

Our results show then that performing four or more symmetry breakings on these 23 special points in moduli space produces every possible nonabelian gauge symmetry in the parent six dimensional theory. This is analogous to the case of discrete gauge symmetries as analysed in \cite{Kachru:2016ttg}. In this reference, however, the lattice $\Gamma_{8,24}$ is taken as the full nonperturbative charge lattice of the theory in three and not two dimensions. The reader is free to interpret our results also in that context. 

We consider the two dimensional case since the gauge groups given by the Niemeier lattices are visible as current algebras in the heterotic string worldsheet CFT. The fact that the Narain lattice splits into two Euclidean lattices signifies that the worldsheet CFT factorizes into two chiral (s)CFTs. The left moving CFT is a meromorphic CFT with $c=24$ with nontrivial current algebras at level 1 while the right moving sCFT is the meromorphic sCFT with $c=12$ based on the $E_8$ lattice. These worldsheet CFTs are examples of the family studied in \cite{Harrison:2021gnp}. These facts will play a key role in understanding the overall structure of the moduli space in six dimensions.

\section{Chaudhuri-Hockney-Lykken Component}\label{s:CHL}
It will be our main objective to extend the analysis carried out so far to other connected components in the moduli space. These take the generic form
\begin{equation}
    \mathcal{M} = O(\Gamma_c) \backslash \text{Gr}(4,20-r)\,,
\end{equation}
where $\Gamma_c$ is the charge lattice analog to the Narain lattice in the standard component, and $r \in \{8,12,14,16,18,20\}$\footnote{For the predicted non-cyclic orbifolds in Section \ref{s:Overall}, we also have $r = 15,17,19$.}. We start by focusing our attention on the unique component with $r = 8$, which we will refer to as the Chaudhuri-Hockney-Lykken (CHL) component.

\subsection{Basic generalizations}

The CHL component \cite{Chaudhuri:1995fk,Chaudhuri:1995bf} can be described in various ways, most notably as an asymmetric orbifold \cite{Narain:1986qm} of the $E_8 \times E_8$ heterotic string on $T^4$ which realizes a holonomy along one of the compact directions whose action exchanges the $E_8$ factors. There are various other descriptions, which can be found in \cite{deBoer:2001wca}. Its charge lattice was computed by Mikhailov \cite{Mikhailov:1998si} and is of the form
\begin{equation}
    \Gamma_c = \Gamma_{1,1}\oplus \Gamma_{3,3}(2) \oplus E_8\,.
\end{equation}
Here the parentheses denotes a scaling of $\Gamma_{3,3}$ by $\sqrt{2}$. This theory exists in nine dimensions, where the charge lattice is $\Gamma_{1,1}\oplus E_8$, and each compactification on $S^1$ extends it by adding $\Gamma_{1,1}(2)$. This can be interpreted as a relative reduction in the size of the circle supporting the holonomy, and is a generic feature of rank reduced charge lattices.

The Mikhailov lattice $\Gamma_c$ enjoys various presentations. In particular, it can be written as
\begin{equation}
    \Gamma_c = \Gamma_{4,4}\oplus [8\, A_1\,|\mathbb{Z}_2]\,,
\end{equation}
where the lattice $[8\, A_1\,|\mathbb{Z}_2]$ is the weight lattice of the gauge group $SU(2)^8/\mathbb{Z}_2$ with $\mathbb{Z}_2$ diagonal. Moreover, $\Gamma_c$ admits a primitive embedding into $\Gamma_{4,20}$, wherein its orthogonal complement is also $[8\, A_1\,|\mathbb{Z}_2]$. This rank 8 lattice can be compactly written as $D_8^*(2)$, but it is the form we have chosen which will generalize to other components.

Importantly, $\Gamma_c$ is not self-dual. This calls for generalizations of the statements made for the Narain lattice regarding symmetry enhancements. Firstly, root lattices may have roots with norm 4 instead of 2, leading to gauge algebras of the type BCF when they are mixed with roots with norm 2, and scaled type A algebras otherwise. Secondly, if a root sublattice has longest root with norm $4$, its current algebra level is 1; if the longest root has norm 2, the level is 2. Finally, the topology is not directly encoded in the overlattice of the root lattice with respect to its embedding into $\Gamma_c$, to which we have referred as $W_G$ in the Narain component. The fundamental group $\pi_1(G)$ can be obtained instead as the quotient $W_G^\vee /  (W_G)_\text{root}^\vee$, where $(W_G)_\text{root}^\vee$ is the coroot lattice and $W_G^\vee$ its overlattice with respect to an embedding into $\Gamma_{c}^*$ \cite{Cvetic:2021sjm}. Indeed note that this procedure reduces to the one outlined in the Narain component in the case that $\Gamma_c$ is self-dual and $G$ is simply-laced. 

We refer the reader to \cite{Font:2021uyw} for more details on the CHL string and its symmetry enhancements.

\subsection{From Narain to CHL: the rank reduction map}\label{ss:nartochl}

The procedure of computing the possible gauge groups from lattice embeddings is rather tedious. Fortunately, there is an alternative method. There exists a map that takes as input some suitable gauge group arising in the Narain component and returns a gauge group in the CHL component \cite{Fraiman:2021hma}. It will be instructive to briefly review its derivation from lattice embedding techniques.

As we have remarked above, $\Gamma_c$ admits a primitive embedding into $\Gamma_{4,20}$ with orthogonal complement $[8\, A_1\,|\mathbb{Z}_2]$. This implies that to any weight lattice $W_G \hookrightarrow \Gamma_{4,20}$ for which $[8\, A_1\,|\mathbb{Z}_2] \hookrightarrow W_G$ one can associate another lattice in $\Gamma_c$ as the orthogonal complement of $[8\, A_1\,|\mathbb{Z}_2]$ in $W_G$. From this newfound lattice one may apply the procedure outlined above to compute the corresponding gauge group $G'$, establishing a map $G \mapsto G'$. That $[8\, A_1\,|\mathbb{Z}_2]$ is primitively embedded into $W_G$ implies that the gauge group $G$ is an enhancement of $SU(2)^8/\mathbb{Z}_2$. The generator $k$ of this $\mathbb{Z}_2$ will correspond then to an order two element in $\pi_1(G)$, from which we can determine that $G$ can be mapped to some $G'$ in the CHL component. 

The effect of taking the orthogonal complement of $[8\, A_1\,|\mathbb{Z}_2]$ in $W_G$ and computing the gauge group $G'$ using the rules for the CHL string, for every one of the possible groups $G$ in the Narain component, makes clear that at the level of the gauge groups the map is as follows. Let 
\begin{equation}
    G = \frac{\tilde G}{\pi_1(G)} = \frac{\tilde G_1 \times \cdots \times \tilde G_s}{\pi_1(G)}\,,
\end{equation}
where $\tilde G$ is the universal cover of $G$ and $\tilde G_i$ its simple factors. Denote by $k = (k_1,...,k_s)$ the elements of $\pi_1(G)$, with $k_i$ the projections of $k$ onto the centers $Z(\tilde G_i)$. Let $\ell$ be an order 2 element of $\pi_1(G)$ corresponding to an element in the lattice $W_G$ which reduces to the order two weight vector in its sublattice $[8\, A_1\,|\mathbb{Z}_2]$. If the projection $\ell_i$ of $\ell$ onto $Z(\tilde G_i)$ is nonzero, $\tilde G_i$ undergoes a transformation according to the rules
\begin{eqnarray}\label{RulesCHL}
    \begin{tabular}{ccc}
			$\tilde G_i$&$\tilde G_i'$&$\ell_i$\\ \hline\hline
			$SU(2)$ & $\emptyset$ & 1\\ \hline
			$SU(2n)$ & $SU(n)$& $n \geq 2$\\ \hline
			$Spin(2n)$ & $Sp(n-2)$ & $v$\\\hline
			$Spin(4n)$ & $Spin(2n+1)$ & $s$ \\\hline
			$E_7$ & $F_4$ & 1\\\hline
		\end{tabular}
\end{eqnarray}
Here $v$ and $s$ are respectively the vector and positive chirality spinor classes in the center of $Spin(2n)$, the later of which is always of order 2 in $Spin(4n)$. Depending on the chosen basis, one may have the negative chirality spinor $c$ instead of $s$. The fundamental group transforms as
\begin{equation}
    \pi_1(G) \mapsto \pi_1(G') \simeq \pi_1(G)/\mathbb{Z}_2\,, ~~~~~ \mathbb{Z}_2 = \{0,\ell\}\,.
\end{equation}

The levels of the worldsheet current algebras will also change generically. All of the transformed simple factors are now associated to root sublattices where the longest root has norm 4, and so they are at level 1. Every spectator term will however change its level from 1 to 2. A simple example is provided by a group of the form $Spin(32)/\mathbb{Z}_2 \times G_4$. Indeed, $Spin(32)/\mathbb{Z}_2$ can be broken to $SU(2)^8/\mathbb{Z}_2$, and is mapped to $Spin(17)$. The result is then the gauge group $Spin(17)\times G_4$ with current algebra $\hat B_{8,1} + \hat G_{4,2}$, with hats denoting affinization, as usual, and rightmost subscripts denoting the level.

Remarkably, this same map arises in studying the moduli space of flat connections for a group $G$ over the torus $T^2$ \cite{Schweigert:1996tg}. As such, it was expected that it would appear naturally in the CHL string in eight dimensions \cite{Lerche:1997rr}. Here we see that a precise realization exists instead in six dimensions, but the reason for this is so far elusive. On the other hand, in the more recent mathematical literature this map was used to rederive the classification of meromorphic CFTs with central charge 24 \cite{2017arXiv170805990H}. We will be able to understand precisely how this method is connected with the structure of the moduli space of six dimensional $\mathcal{N} = (1,1)$ string vacua in the following.

\subsection{Symmetry enhancements from orbit lattices}

As we have seen, every possible gauge symmetry group arising in the Narain component can be obtained as a symmetry breaking of a rank 24 gauge group $G$ with $W_G$ one of the 23 Niemeier lattices. Here we wish to show that an analogous result holds in the CHL component. The basic idea is that applying the rank reduction map to a set of gauge groups obtained from some Niemeier lattice $N_I$ is equivalent to applying the same map to the rank 24 gauge group associated to $N_I$ and then extracting gauge groups $G'$ by symmetry breaking.

Let us illustrate this procedure with an example. Consider the Niemeier lattice $N_\beta$ with root sublattice $D_{16}\oplus E_8$, and associated gauge group $Spin(32)/\mathbb{Z}_2 \times E_8$, and delete four nodes in the $E_8$ as above. In every case, we are left with a group $G$ which can be mapped to the CHL string, which may as well have been obtained by deleting four nodes in the $E_8$ of $Spin(17)\times E_8$. In this case, the operations of rank reduction and node deletion commute. Can this be generalized to node deletions in the $Spin(32)/\mathbb{Z}_2$ factor?

We must consider the cases in which deleting a node from the Dynkin diagram of $Spin(32)/\mathbb{Z}_2$ preserves the fundamental group $\mathbb{Z}_2$. It helps to make explicit the primitive embedding $8\, A_1 \hookrightarrow D_{16}$ of the root sublattice of $[8\, A_1\,|\mathbb{Z}_2]$ into the root sublattice of $W_{Spin(32)/\mathbb{Z}_2}$ in the Dynkin diagram:
\begin{eqnarray}
\begin{aligned}
\begin{tikzpicture}[scale = 1.2]
\draw(0,0)--(7,0);
\draw(0.5,0)--(0.5,0.5);
\draw[fill=Periwinkle](0,0) circle (0.1);
\draw[fill=white](0.5,0) circle (0.1);
\draw[fill=Periwinkle](1,0) circle (0.1);
\draw[fill=white](1.5,0) circle (0.1);
\draw[fill=Periwinkle](2,0) circle (0.1);
\draw[fill=white](2.5,0) circle (0.1);
\draw[fill=Periwinkle](3,0) circle (0.1);
\draw[fill=white](3.5,0) circle (0.1);
\draw[fill=Periwinkle](4,0) circle (0.1);
\draw[fill=white](4.5,0) circle (0.1);
\draw[fill=Periwinkle](5,0) circle (0.1);
\draw[fill=white](5.5,0) circle (0.1);
\draw[fill=Periwinkle](6,0) circle (0.1);
\draw[fill=white](6.5,0) circle (0.1);
\draw[fill=Periwinkle](7,0) circle (0.1);
\draw[fill=white](0.5,0.5) circle (0.1);
\end{tikzpicture}
\end{aligned}
\end{eqnarray}
The order 2 element of the fundamental group corresponds to a vector in the real span of the roots whose nodes are colored. If any of these is deleted, the fundamental group becomes trivial. The white nodes, on the other hand, can be deleted without altering $\pi_1$, thus still allowing the rank reduction map to be applied. Deleting one of them and then applying the rank reduction map is in fact equivalent to mapping first $Spin(32)/\mathbb{Z}_2 \to Spin(17)$ and then appropriately deleting one of the eight nodes. 

In this case, the group $Spin(17)\times E_8$ has a corresponding lattice given by the orthogonal complement of $[8\, A_1\,|\mathbb{Z}_2]$ in $N_\beta$. Following \cite{2017arXiv170805990H}, where lattices of this type have also made an appearance to be explained below, we will refer to it as an orbit lattice (more details are given in Section \ref{ss:hohn}). Indeed, the rank reduction map has an action on the affine Dynkin diagram given by a folding which corresponds to an element of the center of $\tilde G$, and accordingly, the resulting affine algebra is known as an orbit (affine) Lie algebra \cite{Schweigert:1996tg}. 

There are 17 primitive embeddings of $[8\, A_1\,|\mathbb{Z}_2]$ into Niemeier lattices, each of which produces a different orbit lattice for the CHL component. The same considerations as above apply, and so we have 17 special rank 16 gauge groups from which every possible gauge group in the CHL component can be obtained by four or more symmetry breakings. These are recorded in Table \ref{tab:ocyclic}. As in the Narain component, the results obtained in this way agree with those found in \cite{Fraiman:2021hma} using an exploration algorithm.

\subsection{Connection to the 2D theory}\label{ss:CHL2D}
Mirroring the discussion in Section \ref{ss:Narain2D}, we now want to show that the 17 gauge groups corresponding to orbit lattices $N_I'$ in the CHL component are realized in the moduli space of the theory when compactified further on $T^4$. To this end it suffices to prove that there is an isomorphism
\begin{equation}\label{CHLiso}
    \Gamma_{4,4}\oplus \Gamma_{4,4}(2)\oplus [8\, A_1\,|\mathbb{Z}_2] \simeq N_I' \oplus E_8(\text{-}2)\,, 
\end{equation}
for each possible $N_I'$, analogous to the isomorphisms in \eqref{NarainIso}. 

The easiest way to prove eq. \eqref{CHLiso} is by showing that each one of the 18 lattices involved belong to the same lattice genus using for example \texttt{SAGEMATH} or \texttt{MAGMA}. It is a standard theorem of lattice theory that for even lattices with $\Gamma_{1,1}$ sublattice each genus contains exactly one lattice. More generally, it can be proven using standard lattice embedding theorems that any pair of lattices $\Gamma_c$ and $M$ defined as the orthogonal complement of some lattice $\Lambda$ primitively embedded respectively into $\Gamma_{4,20}$ and $N_I$ enjoys the relation
\begin{equation}
    \Gamma_c \oplus \Gamma_{4,4}(n) \simeq M\oplus E_8(\text{-}n)
\end{equation}
for arbitrary $n$; \eqref{CHLiso} is a special case. To see that there are no other possible orthogonal decompositions, it suffices to compute the genus of the Euclidean lattices $N_I'$, which consists exactly of these 17 lattices. Indeed, taking orthogonal complements in the Niemeier lattices with respect to embeddings of a given lattice, in this case $[8\, A_1\,|\mathbb{Z}_2]$, is a standard way of computing lattice genera, known as the Kneser-Nishiyama method \cite{zbMATH00993515 ,zbMATH01236805}. 

From \eqref{CHLiso} we also learn that the 17 gauge groups associated to the orbit lattices arise in points in moduli space where the heterotic worldsheet CFT factorizes. This has the implication that the current algebras at these points correspond to chiral CFTs with $c = 24$. As we will see, for all cyclic orbifolds these are in fact meromorphic; such algebras were notably classified by Schellekens in \cite{Schellekens:1992db} and we find indeed that this set of 17 algebras are reported in this reference. The right-moving sCFT is again the one based on the $E_8$ lattice. 

We have arrived at the notable fact that applying the rank reduction map on the Niemeier lattices has produced current algebras for meromorphic CFTs with $c = 24$. This is in fact a general occurrence, and has been worked out in the (recent) mathematical literature \cite{2017arXiv170805990H}. We will review this work in the next section. It will provide us with enough information to propose a natural partial classification of the components in the moduli space of six dimensional string vacua with $\mathcal{N}=(1,1)$, namely of those defined by orbifolds given by a group of the type $\mathbb{Z}_n$. In each case we are able to extend the methods developed so far to the corresponding moduli space and the gauge symmetries it presents.

\section{Other Components of Cyclic Orbifold Type}\label{s:General}
The results of last section can be generalized straightforwardly to other known components of the moduli space, provided that we know how to extract the gauge group data from a lattice embedding into the charge lattice. Obtaining this information is however quite difficult in general, as we usually require explicit formulas for the masses of the states in the spectrum in some stringy description. 

We addressed this problem in \cite{Fraiman:2021soq} for four components with rank reduced gauge group apart from the CHL string which can be described in seven dimensions by asymmetric orbifolds of the heterotic string on $T^3$, known as heterotic $\mathbb{Z}_n$-triples \cite{deBoer:2001wca}. The rank reduction map was consequently obtained for the corresponding six dimensional theories in \cite{Fraiman:2021hma}. Here too its action on the affine algebras produces an orbit Lie algebra according to the order of an associated element in the fundamental group of the gauge group $G$ in the Narain component. In the following we will see that this map is in perfect agreement with the more general construction proposed in this paper, and so are therefore the derived rules for reading the gauge symmetries from lattice embeddings.

Our approach will be to take an alternative route to the derivation of these rules by focusing on the relation between orbit lattices and meromorphic CFTs with $c = 24$ (and sCFTs with $c = 12$), which has been worked out in the literature, and lifting them from two to six dimensional theories. These results apply in fact to almost every known moduli space component, and predicts the existence of eight more. The outliers correspond to orbifolds not of cyclic type, which will be considered in Section \ref{s:Overall}. The fact that the orbit lattices encode every possible symmetry enhancement in the six dimensional theories ensures that the rules can be derived in complete generality. 

\subsection{Höhn's construction}\label{ss:hohn}

In \cite{Schellekens:1992db} Schellekens notably gave a classification of meromorphic CFTs with $c = 24$, of which 69 have semisimple current algebras. More recently, Höhn showed in \cite{2017arXiv170805990H} that all of these can be obtained in a simple manner from the Niemeier lattices and the possible orbit lattices that can be constructed from them as we have exemplified for the CHL string. Let us explain how this works.

Let $L$ be an even lattice with automorphism group $O(L)$, and $g \in O(L)$ an order $n$ automorphism. The action of $g$ on $L$ leaves invariant a primitive sublattice $L^g$, referred to as the \textit{invariant lattice}. The orthogonal complement of $L^g$ in $L$, denoted by $L_g$,  is in turn referred to as the \textit{coinvariant lattice}. Now suppose that $L$ is a Lie algebra lattice with root sublattice $L_\text{root}$, and \textit{glue code} the abelian group $C_L = L/L_\text{root}$. Each element in $C_L$ defines a special lattice automorphism $g$ corresponding to a symmetry of the affine Dynkin diagram of $L_\text{root}$. From each simple root we can construct a $g$-orbit of roots, their sum being a vector in $L$ invariant under $g$. In the case that two roots have nontrivial inner product, their sum must be multiplied by 2.\footnote{In \cite{2017arXiv170805990H} this rule is not implemented; there is rather a scaling of certain root sublattices. Both procedures allow to get the correct root systems for Schellekens list, but it is the former which makes sense from the string theory point of view, leading to the map of orbit affine Lie algebras of \cite{Fuchs:1995zr,Schweigert:1996tg,Lerche:1997rr}. The corrected results also agree with the current algebras for the $c = 12$ sCFT of \cite{Harrison:2020wxl}.\label{f4}} This yields an \textit{orbit root lattice} $R$, and so one can construct an invariant primitive sublattice by taking its overlattice inside $L$. We refer to the latter as an \textit{orbit lattice}.

We see then that for a given element $g$ in $L/L_\text{root}$ there is a canonical mapping $L \mapsto L^g$. This map can be promoted to one between affine Lie algebras if one is able to consistently assign them to $L$ and $L^g$. For the problem at hand we take $L$ to give a semisimple affine algebra $\mathfrak g$ of ADE type at level 1; equivalently we take all the roots of $L$ to be precisely the norm 2 vectors. For $L^g$, we take the roots to be those vectors obtained from the $g$-orbits of roots, as explained above. The resulting root sublattice $L^g_\text{root}$ is then an orthogonal sum of possibly rescaled root lattices, not necessarily of ADE type. From this we read off the simple algebras in the sum $\mathfrak{g}^g = \mathfrak{g}^g_1 \oplus \mathfrak{g}^g_2 \oplus \cdots $; their levels are taken to be equal to $2m/\alpha_\ell^2$, where $m$ is the order of $g$ times an integer $\lambda$ to be specified later, and $\alpha_\ell$ the longest root in the corresponding root sublattice. 

Take now the set of Niemeier lattices, compute all of their corresponding orbit lattices and assign algebras as above. As Höhn noticed, the subset of algebras with dimension strictly greater than 24 matches precisely those in Schellekens' list if $\lambda$ is appropriately chosen (in most cases $\lambda = 1$ while in a few $\lambda = 2$). We refer the reader to Table 3 of \cite{2017arXiv170805990H} for the full set of relevant data. 

It was also observed in \cite{2017arXiv170805990H} that the orbit lattices are arranged into various lattice genera. For each such genus, there is a unique coinvariant lattice. For example, a set of 17 orbit lattices corresponding to certain order 2 automorphisms of the $N_I$ belong to the same genus and their coinvariant lattice is $[8\, A_1\,|\mathbb{Z}_2]$. This is just what we already found by studying the CHL string in Section \ref{ss:CHL2D}. One immediately suspects that each orbit lattice genus corresponds to a 6D $\mathcal{N}=(1,1)$ moduli space component and, as we will see, this is the case. This correspondence will in turn help elucidate the interpretation of certain points not yet clarified such as the role of the gluing vectors in the orbit lattices\footnote{A gluing vector in a lattice $L$ is any vector which is not a linear combination of roots. This is not to be confused with the notion of gluing vector used in \cite{Schellekens:1992db}. More concretely, the latter are elements of the \textit{gluing code}, which have \textit{rescaled} gluing vectors.}, the meaning of $\lambda$ and the role of orbit lattices with algebras of dimension 24. 
\\\\
\noindent \textbf{Parentheses on coinvariant lattices}

\noindent Before proceeding let us make some observations concerning the coinvariant lattices; we will call them $\Lambda$ here. As we have mentioned, they appear naturally as the orthogonal complements of the charge lattices $\Gamma_c$ in the Narain lattice $\Gamma_{4,20}$. They all share a common feature, namely that removing any node in the associated Dynkin diagram, the glue code $\Lambda/\Lambda_\text{root}$ is reduced. Physically this means that the associated gauge symmetry group (which is indeed realized in the theory) changes its fundamental group when there is any symmetry breaking; e.g.\ $G = SU(2)^8/\mathbb{Z}_2$ breaks to $SU(2)^7$.

The gluing vectors in $\Lambda$ correspond precisely to those elements in the glue code of the host Niemeier lattice which give the automorphism with respect to which $\Lambda$ is coinvariant. For this reason, a classification of all sublattices of the $N_I$ which exhibit the property mentioned above amounts to a classification of orbit lattices with respect to any automorphism of the $N_I$, cyclic or not, by taking orthogonal complements. This point of view allows us to avoid in practice the process of computing the orbit lattices by explicit use of the automorphisms, which can be very cumbersome.   

\subsection{Known components}
\subsubsection{M-Theory on \texorpdfstring{$(\text{K3}\times S^1)/\mathbb{Z}_n$}{(K3 x S1)/Zn}}\label{ss:k3s1}

In the following we wish to examine the moduli space components that are known, and relate them to the orbit lattice construction just explained. 

We start with M-Theory compactified on an orbifold $(\text{K3}\times S^1)/\mathbb{Z}_n$ given by an order $n$ symplectic automorphism of the K3 surface together with an order $n$ shift along the $S^1$. The possibilities are $n = 1,...,8$, with corresponding charge lattices \cite{deBoer:2001wca}
\begin{eqnarray}
    \begin{tabular}{|c|c|c|}
            \hline
			$n$&$\Gamma_c$&$\Gamma_c^\perp \hookrightarrow \Gamma_{4,20}$\\ \hline\hline
			$1$ & $\Gamma_{4,20}$ & $\emptyset$\\ \hline
			$2$ & $\Gamma_{4,4}\oplus [8\, A_1\,|\mathbb{Z}_2]$ & $[8\, A_1\,|\mathbb{Z}_2]$\\ \hline
			$3$ & $\Gamma_{3,3}\oplus \Gamma_{1,1}(3) \oplus A_2 \oplus A_2$ & $[6\, A_2\, |\mathbb{Z}_3]$ \\ \hline
			$4$ & $\Gamma_{3,3}\oplus \Gamma_{1,1}(4) \oplus A_1 \oplus A_1$ & $[2\, A_1 \oplus 4\, A_3\,| \mathbb{Z}_4]$ \\\hline
			$5$ & $\Gamma_{3,3}\oplus \Gamma_{1,1}(5)$ & $[4\, A_4\,| \mathbb{Z}_5]$ \\ \hline
			$6$ & $\Gamma_{3,3}\oplus \Gamma_{1,1}(6)$ & $[2\, A_1 \oplus 2\, A_2 \oplus 2\, A_5\, |\mathbb{Z}_6]$ \\ \hline
			$7$ & $\Gamma_{2,2}\oplus \left(\begin{smallmatrix}\text{-}4&\text{-}1\\\text{-}1&\text{-}2\end{smallmatrix}\right)$ 
			& $[3\, A_6\,|\mathbb{Z}_7]$ \\ \hline
			$8$ & $\Gamma_{2,2}\oplus \left(\begin{smallmatrix}\text{-}4&0\\0&\text{-}2\end{smallmatrix}\right)$ 
			& $[A_1 \oplus A_3 \oplus 2\, A_7\, |\mathbb{Z}_8]$ \\ \hline
		\end{tabular}
\end{eqnarray}
All of the $\Gamma_c$ admit a primitive embedding into the Narain lattice $\Gamma_{4,20}$, with orthogonal complements recorded in the rightmost column. 

As for the CHL string, which is dual to the $n = 2$ case here, the lattices $\Gamma_c^\perp$ embed primitively into some Niemeier lattices and play the role of the coinvariant lattice with respect to certain lattice automorphisms of order $n$. For $n = 2,3,4,5,6$, these theories are dual to the heterotic triples we examined in \cite{Fraiman:2021soq,Fraiman:2021hma}, and an analysis along the lines of Section \ref{s:CHL} carries through. 

The general picture is as follows. The lattices $\Gamma_c$ transform as
\begin{equation}
    \Gamma_c \mapsto \Gamma_c \oplus \Gamma_{4,4}(n)
\end{equation}
when the corresponding theories are further compactified on $T^4$. The resulting lattices enjoy the isomorphisms
\begin{equation}\label{isogen}
    \Gamma_c \oplus \Gamma_{4,4}(n) \simeq N_I^g \oplus E_8(\text{-}n)\,,
\end{equation}
and so there are special points in the moduli space where the worldsheet CFT factorizes, from the point of view of a heterotic string description that we assume to exist (to our knowledge, they have not been constructed in the literature for $n = 7,8$). The left moving CFT at these points has current algebra given by the $N_I^g$ as explained in the previous section, with $\lambda = 1$. 

Since we know the current algebras at these special points, we can identify the gluing vectors in the $N_I^g$ with the massive states sitting in fundamental representations of the gauge group, allowing to compute their topology; this is so because the orthogonal splitting of the charge lattice at these points implies that the projection of $\Gamma_c$ into $N_I^g$ is $N_I^g$, so that there are no other vectors in the lattice to be taken into account for. This reproduces for $n = 2,...,6$ the results we get from applying the rank reduction map of the $\mathbb{Z}_n$-triples to the groups corresponding to the Niemeier lattices, as in Section \ref{s:CHL}. For $n = 7,8$ we are instead able to learn how the gauge groups should be read from lattice embeddings into $\Gamma_c$ (the relevant rules do not depend on the number of spacetime dimensions). As we expect, the resulting rules are a direct generalization from those of \cite{Fraiman:2021soq}, and so the rank reduction map of \cite{Fraiman:2021hma} is trivially extended to these components. 

From these facts it follows that the the gauge groups given by the orbit lattices yield every possible nonabelian gauge group to be found in the parent six dimensional theories upon four or more symmetry breakings, just as for the CHL string. We leave to Appendix \ref{app:symbreak} some details regarding how the fundamental groups of the gauge groups change under such breakings.

\subsubsection{M-Theory on \texorpdfstring{$(T^4 \times S^1)/\mathbb{Z}_n$}{(T4 x S1)/Zn}}

Another family of M-Theory compactifications with $\mathcal{N} = (1,1)$ corresponds to the orbifolds $(T^4 \times S^1)/\mathbb{Z}_n$, with $n = 2,3,4,6$. These are given by an order $n$ symmetry of the $T^4$ which breaks half the supersymmetries together with an order $n$ shift along the $S^1$. The charge lattices and orthogonal complements are 
\begin{eqnarray}\label{t4s1lats}
    \begin{tabular}{|c|c|c|}
            \hline
			$n$&$\Gamma_c$&$\Gamma_c^\perp \hookrightarrow \Gamma_{4,20}$\\ \hline\hline
			$2$ & $\Gamma_{1,1}\oplus \Gamma_{3,3}(2)$ & $[16\, A_1\, |\mathbb{Z}_2^5]$\\ \hline
			$3$ & $\Gamma_{1,1}\oplus \Gamma_{1,1}(3) \oplus A_2(\text{-}1)$ & $[9\, A_2\,|\mathbb{Z}_3^3]$\\ \hline
			$4$ & $\Gamma_{1,1}\oplus \Gamma_{1,1}(4) \oplus 2\, A_1(\text{-}1)$ & $[6\, A_1 \oplus 4\, A_3\,|\mathbb{Z}_2^2\mathbb{Z}_4]$\\ \hline
			$6$ & $\Gamma_{1,1}\oplus \Gamma_{1,1}(6) \oplus A_2(\text{-}2)$ & $[5\, A_1 \oplus 4\, A_2 \oplus A_5\,|\mathbb{Z}_6]$\\\hline
		\end{tabular}
\end{eqnarray}
In this case, however, the lattices $\Gamma_c^\perp$ are coinvariant with respect to automorphisms of the $N_I$ which define orbit lattices with algebras of dimension 24, respectively 
\begin{equation}\label{t4s1alg}
    8\, \hat A_{1,2\lambda}\,,~~3\, \hat A_{2,3\lambda}\,,~~~ \hat A_{3,4\lambda}+3\, \hat A_{1,2\lambda}\,, ~~~ \hat B_{2,3\lambda} + \hat A_{2,3\lambda} + 2\, \hat A_{1,2\lambda}\,.
\end{equation}
These algebras are not in Schellekens' list; as we now explain, they have to be associated to chiral sCFTs with $c = 12$. 

First we must note that the isomorphisms of eq. \eqref{isogen} are still valid, although, interestingly, for each component now there is only one corresponding orbit lattice. This means that at a special point in the moduli space, in some stringy description, the worldsheet CFT factorizes into meromorphic chiral CFTs, with right moving sCFT based on the $E_8$ lattice. Since the algebras with dimension 24 are not in Schellekens list, the only possibility is that the left moving CFT is an sCFT with $c = 12$, i.e. the relevant string theory is Type II. It was shown in \cite{Harrison:2020wxl} that there is such a CFT, named $F_{24}$, which admits various current algebras precisely of dimension 24, namely
\begin{equation}\label{f24alg}
\begin{split}
    &8\, \hat A_{1,2}\,,~~3\, \hat A_{2,3}\,,~~~ \hat A_{3,4}+3\, \hat A_{1,2}\,, ~~~\hat A_{4,5}\,, ~~~ \hat B_{2,3}+\hat G_{2,4}\,,\\ 
    &~~~\hat B_{2,3} + \hat A_{2,3} + 2\, \hat A_{1,2}\,, ~~~ \hat B_{3,5} + \hat A_{1,2}\,, ~~~ \hat C_{3,4} + \hat A_{1,2}\,.
\end{split}
\end{equation}
As we can see, those of \eqref{t4s1alg} are included here if we set $\lambda = 1$.

We can again associate gauge groups to the orbit lattices for the compactifications at hand and show that from symmetry breaking one can obtain every possible gauge group in the six dimensional theories. Take for example the case $n = 2$, which has gauge group $SU(2)^8/\mathbb{Z}_2^7$. The fundamental group is such that breaking any set of four nodes in the Dynkin diagram one always ends up with the gauge group $SU(2)^4/\mathbb{Z}_2^3$, which is the unique maximal enhancement in the theory. For the other three components, the allowed maximal enhancements are of rank 2 and readily written down, and we observe that in general there are very few possible enhancements (see \cite{fp2022} for a list of enhancements.) 

It turns out that Höhn's construction, with the minor modification in Footnote \ref{f4}, produces all of the algebras in \eqref{f24alg}, so that it not only classifies the algebras for meromorphic CFTs with $c = 24$ but also those for those with $c = 12$. Interestingly, however, some of these seem to be degenerate, in the sense that the associated root lattices can be found as sublattices of two inequivalent orbit lattices. The explanation for this is natural from the point of view of string theory, as we now show.

\subsubsection{Theories with discrete theta angle}\label{ss:theta}

It was recently shown in \cite{HP} that certain components in the moduli space of 7D $\mathcal{N} = 1$ theories come in pairs from the point of view of M-Theory on K3 with frozen singularities, predicting the existence of three new moduli space components. The stringy description was provided in \cite{MMHP}, where it was shown that such pairs can be understood as different versions of some string theories distinguished by the presence of a nontrivial discrete theta angle. 

Upon further compactification on $S^1$, the versions without theta angle are dual to the theories described in the previous section with $n = 2,3,4$. The corresponding versions with theta angle turned on have charge lattices
\begin{eqnarray}
    \begin{tabular}{|c|c|c|}
            \hline
			$n$&$\Gamma_c$&$\Gamma_c^\perp \hookrightarrow \Gamma_{4,20}$\\ \hline\hline
			$2$ & $\Gamma_{4,4}(2)$ & $[16\, A_1\,|\mathbb{Z}_2^4]$\\ \hline
			$3$ & $\Gamma_{2,2}(3) \oplus A_2(\text{-}1)$ & $[9\, A_2\,|\mathbb{Z}_3^2]$\\ \hline
			$4$ & $\Gamma_{1,1}(2)\oplus \Gamma_{1,1}(4) \oplus 2\, A_1(\text{-}1)$ & $[6\, A_1 \oplus 4\, A_3\,|\mathbb{Z}_2\mathbb{Z}_4]$\\ \hline
		\end{tabular}
\end{eqnarray}
Note that the lattices $\Gamma_c^\perp$ have the same root sublattices as those for their cousins in \eqref{t4s1lats}, but the gluing vectors are different (see Table \ref{tab:sing} in Appendix \ref{app:lattices} for details). Each one of these again admits a primitive embedding into some Niemeier lattice, such that its orthogonal complement gives an orbit lattice. 

The orbit lattices corresponding to these theories share the root sublattice with their aforementioned cousins up to a scaling factor implying that the corresponding algebras are
\begin{equation}\label{t4s1algtheta}
    8\, \hat A_{1,\lambda}\,,~~~~~~3\, \hat A_{2,\lambda}\,,~~~~~~ \hat A_{3,2\lambda}+3\, \hat A_{1,\lambda}\,.
\end{equation}
We see then that $\lambda$ must be chosen respectively as $2$, $3$ and $2$ in order to match the algebras in \eqref{f24alg}. It is interesting to note that in each case, $\Gamma_c$ is a sublattice of index precisely $\lambda$ of the charge lattice for the theories without theta angles. It would be good to understand better this relation.

As before, we can assign gauge groups to the orbit lattices and obtain from them the six dimensional symmetry enhancements. The difference with respect to those of the previous section is in the gluing vectors; the gauge algebras are exactly the same, but the gauge groups differ in their topologies, i.e. at the level of massive charged states in the spectrum. Indeed this was already observed in \cite{HP} for the seven and eight-dimensional decompactifications of the $n = 2$ cases.

\subsubsection{Dabholkar-Harvey string island}\label{ss:island}
In \cite{Dabholkar:1998kv} it was shown that certain pure supergravity theories, i.e. with no vector multiplets, can be obtained as the low energy limits of string compactifications, and some explicit constructions were provided. This includes the case of six dimensions and $\mathcal{N} = (1,1)$ supersymmetry, which can be obtained from a $\mathbb{Z}_5$-asymmetric orbifold of the Type II string on $T^4$. The only modulus is the dilaton, and so the charge lattice is negative definite, namely
\begin{equation}
    \Gamma_c = A_4(\text{-}1)\,.
\end{equation}

This theory fits as well into the framework of orbit lattices. The orthogonal complement of $\Gamma_c$ in $\Gamma_{4,20}$ is $[5\, A_4\,|\mathbb{Z}_5^2]$, which is the coinvariant lattice corresponding to the orbit lattice with algebra $\hat A_{4,5}$ in \eqref{f24alg}. Deleting four nodes in the Dynkin diagram completely breaks the gauge symmetry, as should be. 

\subsection{New components}
We have seen how every moduli space component of cyclic orbifold type in the literature (to our knowledge) fits into the framework of orbit lattices. There are however ten lattice genera unmatched. In this section we assign to them new moduli space components together with their charge lattices, and discuss possible stringy constructions. We also show that except for special cases analogous to those of Section \ref{ss:theta}, the $T^2$ compactifications of every theory are dual to four dimensional CHL models, which have been classified in \cite{Persson:2015jka}.
\\\\
\noindent \textbf{A rank 8 and a rank 2 theory}

\noindent Out of the ten new components, eight are string islands. Let us then first describe the two which are not. They correspond to the genera D and J of \cite{2017arXiv170805990H}. The coinvariant lattices are respectively
\begin{equation}
    [12\, A_1\,|\mathbb{Z}_2]\,, ~~~~~ [3\, A_1 \oplus 3\, A_5\,|\mathbb{Z}_6]\,,
\end{equation}
which embed primitively into the Narain lattice $\Gamma_{4,20}$ with orthogonal complements
\begin{equation}\label{newlats1}
    \Gamma_{4,4}(2)\oplus D_4\,, ~~~~~ \Gamma_{2,2}(2) \oplus A_2(\text{-}1)\,.
\end{equation}
To match the appropriate algebra levels, we must choose $\lambda = 2$.

Here we should note that upon compactification on $T^2$ down to four dimensions, the corresponding theories will be dual to certain CHL models described in detail in \cite{Persson:2015jka}. These are theories constructed e.g.\ as orbifolds of Type II strings on $\text{K3}\times S^1 \times S^1$, given by an order $n$ symmetry of the K3 sigma model together with an order $2n$ shift along one $S^1$, with $n = 2,6$, respectively. Here we have that $n$ is just the order of the automorphism on the Niemeier lattice which defines the orbit lattices, but the actual order of the orbifold symmetry in the physical theory is $2n$. 

As explained in \cite{Persson:2015jka}, the extra factor of 2 in the shift is necessary for it to be made along a physical compact direction. Indeed, the lattices in \eqref{newlats1} get extended to
\begin{equation}\label{newlats1'}
    \Gamma_{4,4}(2)\oplus \Gamma_{2,2}(4)\oplus D_4\,, ~~~~~ \Gamma_{2,2}(2) \oplus \Gamma_{2,2}(12) \oplus A_2(\text{-}1)
\end{equation}
upon compactification on $T^2$, matching those in the classification of \cite{Persson:2015jka}. Here then the choice of $\lambda$ can be understood as a correction of the overall order of the orbifold symmetry such that a shift can be made along an $S^1$. This is in contrast with the cases in Section \ref{ss:theta}, where $\lambda > 1$ but the shifts are unaffected (the reason for this is not yet clear to us). 

Another peculiarity of these two theories is that they are the only ones of cyclic orbifold type for which there are two orbit lattices that are isometric. It would be interesting to understand the physical meaning of this.  

Since these theories are related to meromorphic CFTs with $c = 24$ it is likely that they can be formulated as asymmetric orbifolds of heterotic strings on $T^4$ or perhaps F-Theory on $(\text{K3} \times T^2)/\mathbb{Z}_n$ with order $2,~6$ symmetries of F-Theory on $\text{K3}\times S^1$ and order $n = 4, 12$ shifts along the remaining $S^1$.
\\\\
\noindent \textbf{Five Type II string islands}

\noindent There are other five genera of orbit lattices associated to current algebras of dimension 24 and rank 4. These then correspond to Type II string islands, similarly to the one described in Section \ref{ss:island}. The charge lattices are (12, 19-21 and 23 in Table \ref{tab:chargecyclic})  
\begin{eqnarray}
 \begin{tabular}{|c|c|c|}
            \hline
$n$&$\Gamma_c$&$\Gamma_c^\perp \hookrightarrow \Gamma_{4,20}$\\ \hline\hline
			$5$ & $\left(\begin{smallmatrix}
   \text{-}4 & 1 & 1 & 1 \\
   1&\text{-}4 & 1 & 1 \\
   1&1&\text{-}4 & 1 \\
   1&1&1&\text{-}4  \\
   \end{smallmatrix}\right)$& $[5\, A_4\,|\mathbb{Z}_5]$\\ \hline
	$8$ & $A_1(\text{-}1) \oplus A_3(\text{-}1)$& $[3\, A_1 \oplus A_3 \oplus 2\, A_7\,|\mathbb{Z}_2\mathbb{Z}_8]$\\ \hline
	$8$ & $3\,A_1(\text{-}1) \oplus A_1(\text{-}2)$& $[3\, A_1 \oplus A_3 \oplus 2\, A_7\,|\mathbb{Z}_8]$\\ \hline
   $10$ & $\left(\begin{smallmatrix}
   \text{-}2 & 0 & 0 & 1 \\
   0&\text{-}2 & 0 & 1 \\
   0&0&\text{-}2 & 1 \\
   1&1&1&\text{-}4  \\
   \end{smallmatrix}\right)$& $[3\, A_1 \oplus 2\,A_4 \oplus  A_9\,|\mathbb{Z}_{10}]$\\ \hline
   	   $12$ & $\left(\begin{smallmatrix}
   \text{-}2 & 1 & 1 & 1 \\
   1&\text{-}2 & 0 & 0 \\
   1&0&\text{-}2 & 0 \\
   1&0&0&\text{-}4  \\
   \end{smallmatrix}\right)$& $[2\, A_1 \oplus 2\,A_2 \oplus  A_3 \oplus A_{11}\,|\mathbb{Z}_{12}]$\\ \hline
		\end{tabular}
\end{eqnarray}

Interestingly, we see that among the string islands there are two pairs corresponding to $n = 5$ and $n = 8$ (cf. Section \ref{ss:island}), in the sense that their coinvariant lattices share root sublattices. This is precisely what happens in the theories that we know that admit discrete theta angles, and so it suggests that the Dabholkar-Harvey string island admits such a theta angle and the second item in the table above also. It would be interesting to explore this further. As before, the three models ``without discrete theta angle" can be identified upon compactification on $T^2$ with CHL models in \cite{Persson:2015jka}.
\\\\
\noindent \textbf{Heterotic island}

\noindent Finally there is a string island associated to the orbit lattice $[2\,A_1 \oplus 2\, A_9 \, | \mathbb{Z}_{10}]$, with charge lattice $D_4(\text{-}1)$. It is associated to a meromorphic CFT with $c = 24$ and so we conjecture that it may be realized for example as an heterotic asymmetric $\mathbb{Z}_{10}$-orbifold.

\subsection{Symmetry enhancements in higher/lower dimensions}

Let us close this section with a discussion of symmetry enhancements in other dimensions deriving from the results we have presented above. 

\subsubsection{The seven dimensional case}\label{ss:7denh}
In \cite{Fraiman:2021soq} we computed the possible symmetry enhancements in seven dimensional theories with heterotic string description, namely the $\mathbb{Z}_n$-triples of \cite{deBoer:2001wca}. There are however other components of the moduli space which can be described e.g.\ by M-Theory on a K3 surface with partially frozen singularities or F-Theory on $(T^4 \times S^1)/\mathbb{Z}_n$ with possible discrete theta angles \cite{HP,MMHP}. The circle compactification of these theories belong to the family of moduli spaces discussed in this paper, and we have in particular learned how to compute their possible symmetry enhancements from the associated lattice embeddings. The rules for reading these gauge groups are independent of the number of spacetime dimensions and so we are now in a position to write down exactly what are the possible symmetry enhancements in the aforementioned seven dimensional moduli spaces.

In Table \ref{tab:otherdim} we record the charge lattices of the seven dimensional theories, together with the possible uplifts to eight and nine dimensions. For F-Theory on $(T^4\times S^1)/\mathbb{Z}_2$ there is only one maximal enhancement, namely $SU(2)^3/\mathbb{Z}_2^2$ with $\pi_1$ generated by the elements $(1,1,0)$ and $(0,1,1)$ of the center $Z(SU(2)^3)$. These are indeed the only rank 3 groups which can be obtained from those in the six dimensional theory by symmetry breaking, hence this result is as expected. In turning on the discrete theta angle, the maximal enhancement is now the simply connected $SU(2)^3$. Both of these theories can be lifted to eight dimensions, wherein the respective maximal enhancements are $SU(2)^2/\mathbb{Z}_2 \simeq SO(4)$ and $SU(2)^2$, in accordance with the results of \cite{HP,Cvetic:2022uuu}. In nine dimensions both of the components have enhancements to $SU(2)$. In all cases discussed so far the level of the current algebra in the worldsheet is 2, in accordance with the results of \cite{HP,MMHP}.

The remaining moduli space components have rank 1 gauge groups. In the case of $(T^4\times S^1)/\mathbb{Z}_3$ there is one $SU(2)$ enhancement at level 3 in both components (i.e. with and without theta angle.) For $(T^4\times S^1)/\mathbb{Z}_4$ there are two $SU(2)$ enhancements in each component, one at level 2 and one at level 4; this is opposed to the results of \cite{HP} where one of the enhancements was naively read off as an $SO(3)$. Finally, there are three $SU(2)$ enhancements in the $(T^4 \times S^1)/\mathbb{Z}_6$ component, at level 2, 3 and 6 respectively. 

These results can be summarized in the statement that for these components with low rank in seven dimensions deleting five nodes in the Dynkin diagrams of the associated orbit lattices one obtains exactly every possible symmetry enhancement. This is manifestly not true for the $\mathbb{Z}_n$-triples; deleting five arbitrary nodes in the corresponding orbit lattices one gets gauge groups which are actually not realized in the theory. The statement above is therefore quite remarkable, and in fact applies to the corresponding theories in eight and nine dimensions.

\subsubsection{The cases $D = 2,3,4,5$}

Having the rules for reading off gauge groups from lattice embeddings we can in principle use the exploration algorithm of \cite{Font:2020rsk,Font:2021uyw,Fraiman:2021soq} to find maximal symmetry enhancements in the compactifications of the theories discussed here. This is however far out of the scope of this paper. We wish instead to discuss a few aspects of symmetry enhancements in these compactified theories in relation to the orbit lattices and their associated gauge groups. 

We start with a result regarding the Narain component, stating that deleting $n \leq 4$ nodes in the Dynkin diagram of some Niemeier lattice $N_I$ one obtains a valid gauge group in $D = 2+n$. To see this, consider first the case $n = 1$. What we want to show is that any primitive sublattice $W$ of an $N_I$ of rank 23 defined by deleting a node in the Dynkin diagram admits a primitive embedding into $\Gamma_{7,23}$. To this end we note that the orthogonal complement of $W$ in $N_I$ can be embedded primitively into $E_8$, since indeed any even lattice of rank $\leq 3$ admits such an embedding. The rank 7 orthogonal complement $T$ in $E_8$ will then have the same discriminant group and quadratic form as $W$, which guarantees the existence of a primitive embedding of $W$ into $\Gamma_{7,23}$ (see e.g.\ Section 3.1 of \cite{Font:2020rsk}.) It is straightforward to extend this result to the cases $n = 2,3$; the case $n = 4$ is of course already a main result of this paper. 

The general situation is that Niemeier lattices encode exactly every possible gauge group realized in the six dimensional theory, in lower dimensions giving a proper subset of them and in higher dimensions giving a proper superset. This is easy to see e.g.\ in nine dimensions where there are only 44 maximal enhancements \cite{Font:2020rsk} while deleting 7 arbitrary nodes off of the Niemeier lattices obviously produces many more. 

This situation is in contrast to what we have already observed in the moduli space components discussed in the previous section. In fact we find it natural to conjecture that theories which are described by worldsheets with $c_L = 12$ in two dimensions are such that their orbit lattices give the only possible maximal enhancement, and its breakings every possible enhancement in the higher dimensional theories obtained by decompactification. This is not too unexpected in light of the fact that in these moduli space components, where a description is available to us, there are no perturbative D-branes \cite{MMHP}. The symmetry enhancements come from nonperturbative effects in a very restricted manner. It would be very interesting to explore this conjecture further in the various stringy descriptions. 

Finally, it would be interesting to determine if the facts we established above for the Narain component hold also for other components with $c_L = 24$ in the worldsheet CFT. The difficulty that we encounter is that the orbit lattices in these cases are not self-dual, and it is not clear how to work with primitive embeddings and orthogonal complements therein.

\section{Extension to Non-Cyclic Orbifolds}\label{s:Overall}
What we have seen so far can be summarized as follows. The rank reduction map (see Table \ref{tab:rrmap}) applied on the current algebras corresponding to the Niemeier lattices produces the classification of current algebras for meromorphic (s)CFTs with $c = 24$ ($c = 12$). These algebras appear naturally in 2D string theory vacua with sixteen supercharges, which can be decompactified to six dimensions to give a list of moduli spaces covering every known case of cyclic orbifold type and predicting eight more. 

A cyclic orbifold symmetry can then be associated to an application of the rank reduction map atOrder the level of the gauge groups, suggesting that non-cyclic orbifolds are related to successive applications. This motivates us to consider orbit lattices $N_I^\mathcal{G}$ embedded into Niemeier lattices $N_I$ which are invariant under non-cyclic subgroups $\mathcal{G}$ of $O(N_I)$. We can then define, as usual, the charge lattice $\Gamma_c$ of the corresponding string theory as the orthogonal complement of $N_I^\mathcal{G}$ in $\Gamma_{4,20}$.

\begin{table}
    \center
    \begin{tabular}{cccc}
		Order	& $\tilde G_i$&$\tilde G_i'$&$\ell_i$\\ \hline\hline
$q$	&		$SU(q)$ & $\emptyset$ & 1\\ \hline
$q$	&		$SU(qn)$ & $SU(n)$& $n \geq 2$\\ \hline
 2&			$Spin(2n)$ & $Sp(n-2)$ & $v$\\\hline
2	&		$Spin(4n)$ & $Spin(2n+1)$ & $s$ \\\hline
2	&		$E_7$ & $F_4$ & 1\\\hline
3  & 			$E_6$ & $G_2$ & 1\\\hline
4   &   		$Spin(4n+2)$ & $Sp(n-1)$ & $1$ \\\hline
2 &	$Spin(2n+1)$ & $Spin(2n-1)$ & $1$ \\\hline
2 & $Sp(2n)$ & $Spin(2n+1)$ & $1$ \\\hline
 2 & $Sp(2n+1)$ & $Sp(n)$ & $1$ \\\hline
		\end{tabular}
	\caption{Rank reduction map for non-simply connected gauge groups. Compare with particular case explained in Section \ref{ss:nartochl}. The level of the associated current algebras will depend on the specific string theory, being related for example to how the root lattices change in scale (cf. Section \ref{ss:hohn}). For $Spin(4n+2)$, we have that $\{0,1,2,3\} \simeq \{0,s,v,c\}$.}\label{tab:rrmap}
\end{table}

From our discussion in \ref{ss:CHL2D} we will still have the isomorphisms
\begin{equation}
    \Gamma_c \oplus \Gamma_{4,4}(n) \simeq N_I^\mathcal{G} \oplus E_8(\text{-}n)\,,
\end{equation}
hence there will always be at least one point in the moduli space of the $T^4$ compactifications where the string worldsheet factorizes. In this cases, however, the left-moving (s)CFTs will not be meromorphic, but rather tensor products of one (s)CFT with $c < 24$ ($c < 12$) with others to complete the appropriate central charge. We have no means to compute the appropriate value of $n$, but we suspect it is generically equal to the order of $\mathcal{G}$ times some constant depending on the specifics of the string theory background, as happens in the cyclic cases \cite{Persson:2015jka}. 

\subsection{Heterotic \texorpdfstring{$\mathbb{Z}_2 \times \mathbb{Z}_2$}{Z2xZ2}-quadruple}\label{ss:z2z2}
Let us illustrate the problem with an example. Compactifying the $\mathbb{Z}_2$-triple from seven to six dimensions one can orbifold the theory again using the $\mathbb{Z}_2$ symmetry that defines the CHL string. This leads to a $\mathbb{Z}_2 \times \mathbb{Z}_2$-quadruple with rank reduction 12 \cite{deBoer:2001wca}, which can be matched with the construction outlined above. 

A complete classification of coinvariant lattices (see Table \ref{tab:sing}) includes two of rank 12, one of which corresponds to a cyclic orbifold. It is therefore natural to assign the other to the theory considered here, namely $[12\, A_1\,|\mathbb{Z}_2^2]$. This is precisely what we need, since this lattice contains inside two copies of $[8\, A_1\,|\mathbb{Z}_2]$, as we would expect from orbifolding two times by a symmetry associated to a rank reduction of 8 (with respect to groups in the Narain component.) The charge lattice is
\begin{equation}\label{quadruplat}
    \Gamma_c = \Gamma_{2,2} \oplus \Gamma_{2,2}(2) \oplus 4\, A_1\,,
\end{equation}
and there are 13 corresponding orbit lattices (see Table \ref{tab:chargenoncyclic}). In this case we have no control over how the gauge symmetry groups are computed in the theory but it is reasonable to suppose that the rules in the cyclic cases generalize. In particular, the gauge groups associated to these 13 orbit lattices can be read from the data in Table \ref{tab:chargenoncyclic} and these should encode all the gauge groups in the six dimensional theory.

To our knowledge, the lattice \eqref{quadruplat} was not computed in the literature (although there is a related lattice in Table 12 of \cite{deBoer:2001wca}.) Note that if we assume that compactifying further on $S^1$ extends the lattice as $\Gamma_c \to \Gamma_c \oplus \Gamma_{1,1}(4)$, the lattice $\Gamma_{c,\text{4D}}$ corresponding to the 4D theory exhibits the symmetry
\begin{equation}
    \Gamma_{c,\text{4D}} = \Gamma_{c,\text{4D}}^*(4)\,,
\end{equation}
meaning that the theory exhibits so-called Fricke S-duality \cite{Persson:2015jka} as we would expect from a heterotic construction.

\subsection{New components}

To our knowledge the only non-cyclic orbifold background in six dimensions is that of the previous section. Apart from this there are 22 more such components suggested by our framework (see Table \ref{tab:chargenoncyclic} in Appendix \ref{app:lattices}). 

It is not clear to us how the corresponding theories could be realized. In many cases we expect that their compactifications to lower dimensions can be constructed without too much difficulty. One could then study the possible decompactification limits of such theories and ensure that the ones we describe do exist. The natural expectation is that this corresponds to taking out a set of $\Gamma_{1,1}(n)$ sublattices from the charge lattices of said lower dimensional theories, such that we recover those listed in Table \ref{tab:chargenoncyclic}.

One possibility is that all of the theories considered here may be obtained as orbifolds of Type IIA string theory on a K3 surface, along the lines of \cite{Persson:2015jka} where cyclic orbifolds of Type IIA on $(\text{K3}\times S^1 \times S^1)$ were classified. This would require the use of many shift vectors along the real span of the lattice $\Gamma_{4,20}$, i.e.\ not on extra circles. It would be interesting to explore this possibility. On the other hand, such theories might also be constructed using Bieberbach manifolds as Type II backgrounds with possible theta angles such as in \cite{MMHP} or heterotic asymmetric orbifolds.

We also note that in the non-cyclic case there are nine string islands, for a total of 16. This significantly extends the results of \cite{Dabholkar:1998kv} where only one was found. A straightforward extension of the methods of this reference is however not enough to find the remaining possibilities. Understanding this problem is also of interest.

In the following we go through some interesting general properties of the proposed moduli spaces and their symmetry enhancements. 

\subsubsection{Uplifting orbifolds of M-Theory on $(\text{K3}\times S^1\times \cdots \times S^1)$}
There are in total fourteen automorphism groups of K3 surfaces with which we may construct M-Theory vacua with 16 supersymmetries. Seven of these are cyclic, of order $2,...,8$, and indeed we have already encountered them in Section \ref{ss:k3s1}. The others have various cyclic subgroups and for each one we require another circle along which we can put a shift such that the orbifold is freely acting. As such, they cannot be constructed in six dimensions. These are (see e.g.\ Sec. 4.2 of \cite{deBoer:2001wca})
\begin{equation}\label{ncaut}
    \mathbb{Z}_2\times \mathbb{Z}_2\,, ~~~~~ \mathbb{Z}_2 \times \mathbb{Z}_4\,, ~~~~~ \mathbb{Z}_2 \times \mathbb{Z}_6\,, ~~~~~ \mathbb{Z}_3 \times \mathbb{Z}_3\,, ~~~~~ \mathbb{Z}_4 \times \mathbb{Z}_4\,, ~~~~~ \mathbb{Z}_2^3\,,~~~~~ \mathbb{Z}_2^4\,,
\end{equation}
of which the first five can be realized in (a maximum of) five dimensions and the last two can be realized in four and three dimensions respectively.

We claim that all of the theories associated to the non-cyclic orbifolds above admit decompactifications up to six dimensions. Indeed we see this is the case for $\mathbb{Z}_2 \times \mathbb{Z}_2$ where the six dimensional theory is just the heterotic quadruple described in Section \ref{ss:z2z2} (it can also be described by F-Theory on $(\text{K3}\times S^1 \times S^1)/(\mathbb{Z}_2 \times \mathbb{Z}_2)$ \cite{deBoer:2001wca}.) To motivate this claim we compare the theories corresponding to the entries 24, 26, 28, 35, 37, 39 and 43 of Table \ref{tab:chargenoncyclic} to the coinvariant sublattices of the lattice $\Gamma_{3,19}$ with respect to the K3 automorphisms,  which were computed in \cite{GS} (see p. 15). We can make an explicit match between these lattices for each $\mathcal{G}$ in \eqref{ncaut} at the level of rank reduction; it is not clear to us how the lattices themselves should be matched. 

Now let us assume that the theories with two generators in $\mathcal{G}$ behave such that $\Gamma_c \to \Gamma_{c}\oplus \Gamma_{1,1}(n)$ with $n = \text{ord}(\mathcal{G})$ upon circle compactification. It can be checked in each case that the resulting charge lattice obeys the relation 
\begin{equation}
    \Gamma_{c,\text{4D}} = \Gamma_{c,\text{4D}}^*(n)\,,
\end{equation}
so that they exhibit as symmetries a form of the Fricke S-dualities discussed in \cite{Persson:2015jka}. This leads us to suspect that these theories can be realized in six dimensions as heterotic quadruples analogous to the one in Section \ref{ss:z2z2}.

The two remaining cases do not exhibit the property above. It is natural instead to propose that they admit non-trivial discrete theta angles in some Type II description. This is easy to see by comparing the lattices 26 and 25 in Table \ref{tab:chargenoncyclic} as well as the lattices 28 and 27. We are therefore led to conjecture that they do not admit heterotic descriptions, as is the case for the theories discussed in Section \ref{ss:theta} and \ref{ss:island}.

\subsubsection{A pair of theories with the same charge lattice}

Now we comment on the theories corresponding to entries 31 and 32 in Table \ref{tab:chargenoncyclic}. They have the same charge lattice
\begin{equation}
    \Gamma_c = \Gamma_{2,2}(2) \oplus 2\, A_1(\text{-}1)\,,
\end{equation}
but these are defined as the orthogonal complement of two different lattices in $\Gamma_{4,20}$, which therefore belong to the same genus. Both of these lattices are of the form $[18\, A_1 \, | \mathbb{Z}_2^6]$, but differ in their gluing vectors (see Table \ref{tab:sing}). In fact, they can be shown to be isometric; the genus, whose elements are isometry classes of lattices, has only one member. 

One qualitative distinction between these two theories is that one has only one associated orbit lattice while the other, which we distinguish with a prime, has two (see Table \ref{tab:ononcyclic}). In both cases there is a maximal enhancement to $SU(2)^2/\mathbb{Z}_2^2 \simeq PSO(4)$, but the latter also admits the simply connected $SU(2)^2 \simeq Spin(4)$.

This situation is similar to that of the $\mathbb{Z}_5$ and $\mathbb{Z}_6$ heterotic triples in seven dimensions, which have the same charge lattice $\Gamma_{3,3}$. They are distinguished of course by the order of the automorphism, but this is not immediately visible at the level of lattices. We see this distinction explicitly in lower dimensions, where they get extended respectively by $\Gamma_{n,n}(5)$ and $\Gamma_{n,n}(6)$. It is possible that for the theories considered here a similar situation occurs. Even though they are associated to the same automorphism group $\mathcal{G} = \mathbb{Z}_2^2$, there could be other effects that change how the lattices transform upon compactification, as we have seen already.
\subsubsection{Appearance of $SO(3)$'s}
A remarkable feature of all the moduli spaces found in the literature is that the gauge symmetry group $SO(3)$ is not realized in them. We know in the case of eight dimensional theories that it is indeed ruled out by Swampland considerations \cite{Montero:2020icj}. In seven dimensions, although there is no analogous constraint, it is the case that this group does not appear (see \cite{Fraiman:2021soq} for the $\mathbb{Z}_n$ heterotic triples and Section \ref{ss:7denh} for the other components.) However, in the potential components corresponding to the entries 25 to 33 in Table \ref{tab:chargenoncyclic} they do appear. This is significant as it allows for there to be possible odd rank reductions. Indeed, any time a gauge group can be broken to $SO(3)$ one can apply the rank reduction map associated to the nontrivial element in $\pi_1(SO(3)) = \mathbb{Z}_2$ reducing the rank by 1. 

\subsubsection{Rank reduction patterns}

If we accept that the theories predicted by our framework do exist, we find an interesting pattern in the allowed ranks of the gauge groups of theories with 16 supercharges. Namely, in nine and eight dimensions the allowed rank reductions are
\begin{equation}
    \text{Rank reduction in 8D, 9D} = 8\,,\,16\,.
\end{equation}
In seven dimensions we find more possibilities,
\begin{equation}
    \text{Rank reduction in 7D} = 8\,,\,12\,,\,14\,,\,16\,,\,18\,.
\end{equation}
The pattern that we want to highlight is that between the possible rank reductions there is a gap, which in eight and nine dimensions is always 8. In seven dimensions, there are three gaps: 8, 4 and 2. Now, if we look at the six dimensional case we get
\begin{equation}\label{rr6}
    \text{Rank reduction in 6D} = 8\,,\,12\,,\,14\,,15\,,...\,,20\,,
\end{equation}
which clearly fits into a pattern; the gaps are 8, 4, 2 and 1, and in particular, odd rank reductions are now possible. This pattern would be perfect if there existed a theory in eight dimensions with rank $18-12 = 6$, such as a decompactification of the heterotic $\mathbb{Z}_3$-triple. However, all evidence suggests that it does not exist (in particular since string universality has already been claimed to hold in eight dimensions \cite{Hamada:2021bbz,Bedroya:2021fbu}.) In any case, it would be interesting to understand this pattern in the rank reductions.

\section{Conclusions}\label{s:Conclusions}
In this paper we have shown that every known connected component in the moduli space of six dimensional string vacua is connected to the Narain component at the level of gauge symmetry groups through a rank reduction map, such that the full scope of application of this map suggests the existence of many other components without any known string theoretical description.

The cases in which the rank reduction map is applied once correspond to cyclic orbifolds, and can be related through compactification on $T^4$ to meromorphic (s)CFTs with $c = 24$ ($c = 12$) appearing in the left moving part of the string worldsheet. We used this connection to determine how nonabelian gauge symmetries are read from lattice embeddings into the charge lattices, and showed that the gauge groups corresponding to the aforementioned meromorphic CFTs encode every possible gauge group in the parent six dimensional theories through symmetry breaking. In this way we have in particular verified the results of \cite{Font:2020rsk,Font:2021uyw,Fraiman:2021soq,Fraiman:2021hma} which were obtained by different constructive means. Among the identified components we find the compactifications of the new theories proposed in \cite{HP,MMHP}, validating such results.

We have essentially suggested that the rank reduction map is always an allowed operation in quantum gravity theories in the regime of six dimensions and $\mathcal{N} = (1,1)$ supersymmetry. As such, consecutive applications of it should be associated to string theory backgrounds given by non-cyclic orbifolds. The only example of such a construction known to us (the heterotic $\mathbb{Z}_2 \times \mathbb{Z}_2$-quadruple of \cite{deBoer:2001wca}) indeed fits into this framework. 

Even though we have interpreted our results as intrinsic to six dimensional theories, one can instead focus in the connection between the various left moving CFTs that appear in the 2D theories. The precise idea is that the meromorphic CFTs given by the Niemeier lattices can be transformed into all the other meromorphic CFTs with nontrivial current algebra such that tensoring with the right moving sCFT based on the $E_8$ lattice the resulting theory can always be decompactified to six dimensions due to having a charge sublattice of the form $\Gamma_{4,4}(n)$. More general transformations lead to other CFTs with $c = 24$ but which are not meromorphic but rather the product of more than one CFT, but these also admit decompactifications to 6D corresponding to non-cyclic orbifolds. These statements hinge on the fact that the right moving sCFT is based on the $E_8$ lattice, which is intimately related to $\Gamma_{4,4}$ and so to decompactifications to six dimensions. It would be certainly interesting to explore the possibility of tensoring chiral CFTs so as to obtain theories that decompactify instead to five dimensions, but this does not seem likely; in other words, such a theory which is intrinsically defined in five dimensions (such as the 5D string island of \cite{Dabholkar:1998kv}) may have an associated 2D compactification whose moduli space does not have points in which the worldsheet CFT factorizes. 

Given that the organizing principle we have proposed for the theories in this paper can be formulated purely at the level of gauge groups, it seems to us that it may not be too hard to derive using Swampland constraints. Together with explicit constructions for every proposed string compactification this would establish string universality in 6D with 16 supercharges. The strategies developed in \cite{Bedroya:2021fbu,Hamada:2021bbz} may be useful in this regard. 

It would also be very interesting to see if the techniques used in this paper can be extended to settings with less supersymmetry, such as M-Theory on certain $\mathbb{Z}_2$ orbifolds of K3 surfaces recently studied in \cite{Acharya:2022shu}; the language of lattices is certainly very useful here. On the other hand, many of the theories we consider here can be formulated also in terms of frozen singularities or Type IIB strings compactified on Bieberbach manifolds, which also have received attention in the 5D $\mathcal{N} = 1$ regime \cite{Cheng:2022nso}. 

Finally, we should also note that when breaking supersymmetry further it may be necessary to consider non-Abelian orbifolds (see e.g. \cite{Fischer:2012qj,Fischer:2013qza}). It would be very interesting to see how these could fit into a picture of the type advocated here. 
\vspace{-0.1in}
\subsection*{Acknowledgements}
We are grateful to Andreas Braun and Peng Cheng for useful discussions. We thank Mariana Graña and Miguel Montero for helpful comments on the manuscript. HPF thanks M. Montero for collaboration on related topics. This work was partially supported by the ERC Consolidator Grant 772408-Stringlandscape, PIP-CONICET-11220150100559CO, UBACyT and ANPCyT-PICT-2016-1358.

\newpage
\appendix

\section{Aspects of Lattices}
In this appendix we record some facts regarding the theory of lattices and lattice embeddings that we use in the main text. We record Niemeier lattices, orbit lattices and some theorems.

\subsection{Niemeier lattices}\label{app:niemeier}

Even unimodular (self-dual) Euclidean lattices exist in dimensions $d \in 8\mathbb{Z}$. For $d = 1$ there is only the $E_8$ lattice; for $d = 2$ there is the lattice $2\, E_8$ and also $W_{Spin(32)/\mathbb{Z}_2}$, obtained by adding the vector $(\frac12^{16})$ to $D_{16}$. For $d = 24$, the latter construction of adding \textit{gluing vectors} to ADE lattices yields 23 different unimodular lattices known as the Niemeier lattices; there is a 24th unimodular lattice which has no roots, known as the Leech lattice.
\begin{table}[H]
	\centering
		\def\arraystretch{1.1}
		\begin{minipage}[t]{0.49\textwidth}\centering
		\strut\vspace*{-\baselineskip}\newline	
  \begin{tabular}{|>{$}c<{$}||>{$}c<{$}|>{$}c<{$}>{\scriptsize$}c<{$}|}\hline
			I& (N_I)_{\text{root}} & \multicolumn{2}{|c|}{$\frac{N_I}{(N_I)_{\text{root}}}$} \\ \hline\hline
			\alpha  & \text{D}_{24} & \mathbb{Z}_2 & 
			\begin{array}{@{}c@{}}
				s \\
			\end{array}
			\\\hline
			\beta  & \text{D}_{16}\oplus\text{E}_8 & \mathbb{Z}_2 & 
			\begin{array}{@{}c@{}c@{}}
				s & 0 \\
			\end{array}
			\\\hline
			\gamma  & 3 \text{E}_8 & 1 & 
			\begin{array}{@{}c@{}c@{}c@{}}
				0 & 0 & 0 \\
			\end{array}
			\\\hline
			\delta  & \text{A}_{24} & \mathbb{Z}_5 & 
			\begin{array}{@{}c@{}}
				5 \\
			\end{array}
			\\\hline
			\varepsilon  & 2 \text{D}_{12} & \mathbb{Z}_2^2& 
			\begin{array}{@{}c@{}c@{}}
				s & v \\
				c & c \\
			\end{array}
			\\\hline
			\zeta  & \text{A}_{17}\oplus\text{E}_7 & \mathbb{Z}_6 & 
			\begin{array}{@{}c@{}c@{}}
				3 & 1 \\
			\end{array}
			\\\hline
			\eta  & \text{D}_{10}\oplus2 \text{E}_7 & \mathbb{Z}_2^2& 
			\begin{array}{@{}c@{}c@{}c@{}}
				s & 1 & 0 \\
				c & 0 & 1 \\
			\end{array}
			\\\hline
			\theta  & \text{A}_{15}\oplus\text{D}_9 & \mathbb{Z}_8 & 
			\begin{array}{@{}c@{}c@{}}
				2 & 1 \\
			\end{array}
			\\\hline
			\iota  & 3 \text{D}_8 & \mathbb{Z}_2^3& 
			\begin{array}{@{}c@{}c@{}c@{}}
		0 & c & c \\
		s & s & s \\
		c & 0 & c \\
			\end{array}
			\\\hline
			\kappa  & 2 \text{A}_{12} & \mathbb{Z}_{13}& 
			\begin{array}{@{}c@{}c@{}}
				1 & 5 \\
			\end{array}
			\\\hline
			\lambda  & \text{A}_{11}\oplus\text{D}_7\oplus\text{E}_6 & \mathbb{Z}_{12}& 
			\begin{array}{@{}c@{}c@{}c@{}}
				1 & 1 & 1 \\
			\end{array}
			\\\hline
			\mu  & 4 \text{E}_6 & \mathbb{Z}_3^2& 
			\begin{array}{@{}c@{}c@{}c@{}c@{}}
		0 & 1 & 1 & 1 \\
		1 & 0 & 1 & 2 \\
			\end{array}
			\\\hline
			\nu  & 2 \text{A}_9\oplus\text{D}_6 & \mathbb{Z}_2 \mathbb{Z}_{10}& 
			\begin{array}{@{}c@{}c@{}c@{}}
				5 & 0 & s \\
				2 & 9 & c \\
			\end{array}
			\\\hline
			\xi  & 4 \text{D}_6 & \mathbb{Z}_2^4& 
			\begin{array}{@{}c@{}c@{}c@{}c@{}}
	0 & s & v & c \\
	0 & c & s & v \\
	s & 0 & c & v \\
	c & 0 & v & s \\
			\end{array}
			\\\hline
			o & 3 \text{A}_8 & \mathbb{Z}_3 \mathbb{Z}_9& 
			\begin{array}{@{}c@{}c@{}c@{}}
				0 & 3 & 6 \\
				1 & 1 & 4 \\
			\end{array}
			\\\hline
			\pi  & 2 \text{A}_7\oplus2 \text{D}_5 & \mathbb{Z}_4 \mathbb{Z}_8& 
			\begin{array}{@{}c@{}c@{}c@{}c@{}}
				0 & 2 & 3 & 1 \\
				1 & 1 & 1 & 2 \\
			\end{array}
			\\\hline
			\rho  & 4 \text{A}_6 & \mathbb{Z}_7^2& 
			\begin{array}{@{}c@{}c@{}c@{}c@{}}
				0 & 1 & 2 & 4 \\
				1 & 0 & 4 & 5 \\
			\end{array}
			\\\hline
		\end{tabular}

		\end{minipage}
		\begin{minipage}[t]{0.5\textwidth}\centering
		\strut\vspace*{-\baselineskip}\newline
		\begin{tabular}{|>{$}c<{$}||>{$}c<{$}|>{$}c<{$}>{\scriptsize$}c<{$}|}\hline
			I& (N_I)_{\text{root}} & \multicolumn{2}{|c|}{$\frac{N_I}{(N_I)_{\text{root}}}$} \\ \hline\hline
			\sigma  & 4 \text{A}_5\oplus\text{D}_4 &\mathbb{Z}_2 \mathbb{Z}_6^2 & 
			\begin{array}{@{}c@{}c@{}c@{}c@{}c@{}}
				3 & 0 & 0 & 3 & c \\
				0 & 2 & 5 & 5 & s \\
				2 & 1 & 0 & 5 & v \\
			\end{array}
			\\\hline
			\tau  & 6 \text{D}_4 & \mathbb{Z}_2^6& 
			\begin{array}{@{}c@{}c@{}c@{}c@{}c@{}c@{}}
			 0 & 0 & s & s & c & v \\
			0 & 0 & c & v & v & c \\
			0 & s & 0 & v & s & v \\
			0 & c & 0 & s & v & s \\
			s & 0 & 0 & v & c & s \\
			c & 0 & 0 & c & v & v \\
			\end{array}
			\\\hline
			\upsilon  & 6 \text{A}_4 &\mathbb{Z}_5^3& 
			\begin{array}{@{}c@{}c@{}c@{}c@{}c@{}c@{}}
 0 & 0 & 1 & 2 & 3 & 4 \\
 0 & 1 & 0 & 4 & 3 & 2 \\
 1 & 0 & 0 & 2 & 1 & 2 \\
			\end{array}
			\\\hline
			\varphi  & 8 \text{A}_3 &\mathbb{Z}_4^4& 
			\begin{array}{@{}c@{}c@{}c@{}c@{}c@{}c@{}c@{}c@{}}
 0 & 0 & 0 & 1 & 1 & 3 & 2 & 1 \\
 0 & 0 & 1 & 0 & 2 & 3 & 3 & 3 \\
 0 & 1 & 0 & 0 & 1 & 1 & 3 & 2 \\
 1 & 0 & 0 & 0 & 1 & 2 & 1 & 3 \\
			\end{array}
			\\\hline
			\chi  & 12 \text{A}_2 &\mathbb{Z}_3^6& 
			\begin{array}{@{}c@{}c@{}c@{}c@{}c@{}c@{}c@{}c@{}c@{}c@{}c@{}c@{}}
 0 & 0 & 0 & 0 & 0 & 1 & 2 & 2 & 2 & 1 & 0 & 1 \\
 0 & 0 & 0 & 0 & 1 & 0 & 1 & 1 & 0 & 1 & 1 & 1 \\
 0 & 0 & 0 & 1 & 0 & 0 & 2 & 1 & 2 & 0 & 1 & 2 \\
 0 & 0 & 1 & 0 & 0 & 0 & 0 & 1 & 2 & 2 & 2 & 1 \\
 0 & 1 & 0 & 0 & 0 & 0 & 1 & 2 & 2 & 2 & 1 & 0 \\
 1 & 0 & 0 & 0 & 0 & 0 & 1 & 0 & 2 & 1 & 2 & 2 \\
\end{array}
			\\\hline
			\psi  & 24 \text{A}_1 &\mathbb{Z}_2^{12}& 
\begin{array}{@{}c@{}c@{}c@{}c@{}c@{}c@{}c@{}c@{}c@{}c@{}c@{}c@{}c@{}c@{}c@{}c@{}c@{}c@{}c@{}c@{}c@{}c@{}c@{}c@{}}
 0 & 0 & 0 & 0 & 0 & 0 & 0 & 0 & 0 & 0 & 0 & 1 & 1 & 1 & 1 & 1 & 0 & 0 & 1 & 0 & 0 & 1 & 0 & 1 \\
 0 & 0 & 0 & 0 & 0 & 0 & 0 & 0 & 0 & 0 & 1 & 0 & 0 & 0 & 0 & 1 & 0 & 1 & 1 & 0 & 1 & 1 & 1 & 1 \\
 0 & 0 & 0 & 0 & 0 & 0 & 0 & 0 & 0 & 1 & 0 & 0 & 0 & 0 & 1 & 0 & 1 & 1 & 0 & 1 & 1 & 1 & 1 & 0 \\
 0 & 0 & 0 & 0 & 0 & 0 & 0 & 0 & 1 & 0 & 0 & 0 & 0 & 1 & 0 & 1 & 1 & 0 & 1 & 1 & 1 & 1 & 0 & 0 \\
 0 & 0 & 0 & 0 & 0 & 0 & 0 & 1 & 0 & 0 & 0 & 0 & 1 & 0 & 1 & 1 & 0 & 1 & 1 & 1 & 1 & 0 & 0 & 0 \\
 0 & 0 & 0 & 0 & 0 & 0 & 1 & 0 & 0 & 0 & 0 & 0 & 1 & 0 & 0 & 1 & 1 & 1 & 0 & 1 & 0 & 1 & 0 & 1 \\
 0 & 0 & 0 & 0 & 0 & 1 & 0 & 0 & 0 & 0 & 0 & 0 & 1 & 1 & 0 & 0 & 1 & 0 & 0 & 0 & 1 & 1 & 1 & 1 \\
 0 & 0 & 0 & 0 & 1 & 0 & 0 & 0 & 0 & 0 & 0 & 0 & 0 & 1 & 1 & 0 & 0 & 0 & 1 & 1 & 1 & 0 & 1 & 1 \\
 0 & 0 & 0 & 1 & 0 & 0 & 0 & 0 & 0 & 0 & 0 & 0 & 1 & 1 & 0 & 0 & 0 & 1 & 1 & 1 & 0 & 1 & 1 & 0 \\
 0 & 0 & 1 & 0 & 0 & 0 & 0 & 0 & 0 & 0 & 0 & 0 & 0 & 1 & 1 & 1 & 1 & 1 & 0 & 0 & 1 & 0 & 0 & 1 \\
 0 & 1 & 0 & 0 & 0 & 0 & 0 & 0 & 0 & 0 & 0 & 0 & 1 & 1 & 1 & 1 & 1 & 0 & 0 & 1 & 0 & 0 & 1 & 0 \\
 1 & 0 & 0 & 0 & 0 & 0 & 0 & 0 & 0 & 0 & 0 & 0 & 1 & 0 & 1 & 0 & 1 & 1 & 1 & 0 & 0 & 0 & 1 & 1 \\
\end{array}
			
			\\\hline
		\end{tabular}
		\end{minipage}
	\caption{Niemeier lattices}\label{tab:Ni}
\end{table}
The Niemeier lattices are listed in Table \ref{tab:Ni}. They are specified by their root sublattice $(N_I)_\text{root}$ together with the gluing vectors, which are encoded in the abelian group $N_I/(N_I)_\text{root}$. This group is specified as a subgroup of the discriminant group of the root sublattice,  $N_I^*/(N_I)_\text{root}$, by a set of generating elements.

\subsection{Correspondence between Narain and Niemeier lattices}\label{app:NarNie}

Here we record some important results regarding a relation between primitive embeddings of lattices into the Narain lattice $\Gamma_{4,20}$ and the Niemeier lattices $N_I$, derived in \cite{Cheng:2016org,Gaberdiel:2011fg}. We have the two following theorems:
\\\\
\noindent \textbf{Theorem 1.} (part of Theorem 1 in \cite{Cheng:2016org}) \textit{Let $G$ be a subgroup of $O^+(\Gamma^{4,20})$ fixing pointwise a sublattice $\Gamma^G$ of signature $(-^4,+^d)$, $d \geq 0$. Then there exists a primitive embedding $i$ of the orthogonal complement $\Gamma_G$ into some positive-definite rank 24 even unimodular lattice $N$ }
\begin{equation}
    i~:~\Gamma_G \hookrightarrow N\,.
\end{equation}

\noindent \textbf{Theorem 2.} (part of Theorem 2 in \cite{Cheng:2016org}) \textit{Let $N$ be a positive definite rank 24 even unimodular lattice and $\hat G$ be a subgroup of $O(N)$ fixing pointwise a sublattice $N^{\hat G}$ of rank $4+d,~d\geq 0$. Then, there exists a primitive embedding}
\begin{equation}
    f~:~ N_{\hat G} \hookrightarrow \Gamma_{4,20}
\end{equation}
\textit{of the coinvariant sublattice $N_{\hat G}$ into the Narain lattice $\Gamma_{4,20}$.}
\\\\
Note that, as opposed to \cite{Cheng:2016org}, we use the conventions in which $\Gamma_{d,d+16}$ has signature $(-^d,+^{d+16})$ and not $(+^d,-^{d+16})$. We are interested in the case in which $\Gamma_G$ has roots and $N$ is a Niemeier lattice $N_I$ (recall that in our conventions we take the Niemeier lattices to be those 23 with roots, separately from the Leech lattice.) 

For our purposes we take $d = 0$ in Theorem 1, so that $\Gamma_G$ is a rank 20 positive definite lattice. The negative definite 4-plane is polarized according to the values of the moduli in the theory, so that it being fixed under a subgroup of $O^+(\Gamma_{4,20})$ means that the corresponding point in moduli space is fixed. If $\Gamma_G$ is a Lie algebra lattice, the corresponding moduli are completely fixed by the T-duality subgroup isomorphic to the Weyl group of $\Gamma_G$. Theorem 1 then states that $\Gamma_G$ can be primitively embedded into some $N_I$. Theorem 2 works inversely, and in general, for $d \geq 0$, we find that any Lie algebra lattice $L$ which admits a primitive embedding into $\Gamma_{4,20}$ also admits one into some $N_I$, and vice versa.   

\subsection{Comments on gauge group topology and breakings}\label{app:symbreak}
Here we make some comments regarding the computation of the fundamental groups of the gauge groups $G$ associated to orbit lattices and their breakings. As discussed int he text, these can be extracted from the gluing vectors therein, which represent the massive states in the spectrum sitting in the allowed representations of $G$. The actual computation is as follows \cite{Cvetic:2021sjm}. Let $W$ be the orbit lattice in question and $L$ its root sublattice. Scale every root as
\begin{equation}
    \alpha \to \alpha' =  \frac{2}{\alpha^2}\alpha
\end{equation}
to obtain the coroot lattice $L^\vee$. Embed $L^\vee$ into the dual charge lattice $\Gamma_c^*(n) \simeq W^*(n)\oplus E_8(-1)$, with $n$ an appropriate scaling, and compute its overlattice $W^\vee$; the quotient $W^\vee/L^\vee$ is then isomorphic to $\pi_1(G)$. We make the observation that this quotient coincides with the so-called glue code of $W$, which can be checked explicitly from looking at Tables 5 to 16 in \cite{2017arXiv170805990H}. To see what happens to $\pi_1(G)$ when a node in the Dynkin diagram is deleted, just select the coroot system corresponding to the remaining nodes and compute the overlattice quotient as above. 

As a simple example consider the lattice $W$ with root sublattice $E_8\oplus B_8$ (note that $B_8 \simeq 8A_1$) and gluing vector $k = (0,1)$. This lattice vector gives rise to massive states in the fundamental representation of $Spin(17)$ and so $G = E_8 \times Spin(17)$, coinciding with the fact that the glue code of $W$ is trivial. Since the associated coroot lattice has no overlattice in $\Gamma_c^*(2)$, any symmetry breaking leaves the gauge group simply connected, similarly to what happens in the Narain component. 

\section{Orbit lattices, charge lattices and coinvariant lattices}\label{app:lattices}
Here we record first the orbit lattices resulting from the construction in Section \ref{ss:hohn}. We use an unified notation $O_A$ in order to make easy reference to it. In each case the orbit lattice is the invariant sublattice of some Niemeier lattice $N_I$ under an automorphism subgroup $\mathcal{G} \subset O(N_I)$, namely $N_I^{\mathcal{G}}$, with coinvariant lattice $N_{I;\mathcal{G}}$ also specified. These lattices are equipped with a choice of root sublattice which is in general not simply-laced, i.e.\ of ADE type. Therefore the group $O_A/(O_A)_\text{root}$ does not agree with the glue code in general. The former specifies the orbit lattice itself while the latter specifies the fundamental group $\pi_1(G)$ of the gauge symmetry group associated to $O_A$, as realized for example in a 2D theory (see previous section.) There are 59 orbit lattices of cyclic type which we list in Table \ref{tab:ocyclic} and 57 of non-cyclic type in Table \ref{tab:ononcyclic}. There are also orbit lattices of rank less than four which we do not include as they are not relevant for six dimensional theories. 

Taking the orthogonal complement of the coinvariant lattices inside $\Gamma_{4,20}$ we obtain the charge lattice $\Gamma_c$ associated to some moduli space component. There are 23 charge lattices of cyclic type, listed in Table \ref{tab:chargecyclic} and 23 of non-cyclic type listed in \ref{tab:chargenoncyclic}. The coinvariant lattices themselves are specified in Table \ref{tab:sing}. We also record the charge lattices of higher dimensional theories in Table \ref{tab:otherdim}.

The notation in these tables is slightly altered for reasons of space and clarity. In particular, scalings such as $A_n(n)$ are written $\text{A}_n^{(n)}$.

\begin{table}[H]
	\centering
		\def\arraystretch{1.1}
\begin{tabular}
  {|>{$}c<{$}||>{$}c<{$}|>{$}c<{$}>{\scriptsize$}c<{$}|>{$}c<{$}>{\scriptsize$}c<{$}|>{$}c<{$}|>{$}c<{$}|}\hline A & (O_A)_{\text{root}} & \multicolumn{2}{|c|}{$\frac{O_A}{(O_A)_{\text{root}}}$}&
   \multicolumn{2}{|c|}{\text{Glue code}}& I & N_{I,\mathcal{G}}    \\ \hline

\hline 1 & \text{B}_8^{\text{}}+\text{E}_8^{\text{}} & \mathbb{Z}_2  & 
\begin{array}{@{}c@{}c@{}}
 1 & 0 \\
\end{array}
 & 1 & 
\begin{array}{@{}c}
 \text{} \\
\end{array}
 & \beta  &
 \\
 \cline{1-7} 2 & \text{B}_6^{\text{}}\oplus\text{C}_{10}^{\text{}} & \mathbb{Z}_2  & 
\begin{array}{@{}c@{}c@{}}
 1 & 1 \\
\end{array}
 & \mathbb{Z}_2  & 
\begin{array}{@{}c@{}c@{}}
 1 & 1 \\
\end{array}
 & \varepsilon  & 
 \\
 \cline{1-7} 3 & \text{C}_8^{\text{}}\oplus2 \text{F}_4^{\text{}} & 1 & 
\begin{array}{@{}c}
 \text{} \\
\end{array}
 & \mathbb{Z}_2  & 
\begin{array}{@{}c@{}c@{}c@{}}
 1 & 0 & 0 \\
\end{array}
 & \eta  & 
 \\
 \cline{1-7} 4 & \text{B}_5^{\text{}}\oplus\text{E}_7^{\text{}}\oplus\text{F}_4^{\text{}} & \mathbb{Z}_2  & 
\begin{array}{@{}c@{}c@{}c@{}}
 1 & 1 & 0 \\
\end{array}
 & \mathbb{Z}_2  & 
\begin{array}{@{}c@{}c@{}c@{}}
 1 & 1 & 0 \\
\end{array}
 & \eta  & 
 \\
 \cline{1-7} 5 & \text{A}_7^{\text{(2)}}\oplus\text{D}_9^{\text{}} & \mathbb{Z}_8  & 
\begin{array}{@{}c@{}c@{}}
 1 & 1 \\
\end{array}
 & \mathbb{Z}_4  & 
\begin{array}{@{}c@{}c@{}}
 2 & 1 \\
\end{array}
 & \theta  & 
 \\
 \cline{1-7} 6 & \text{B}_4^{\text{}}\oplus2 \text{C}_6^{\text{}} & \mathbb{Z}_2  & 
\begin{array}{@{}c@{}c@{}c@{}}
 1 & 1 & 1 \\
\end{array}
 & \mathbb{Z}_2^2  & 
\begin{array}{@{}c@{}c@{}c@{}}
 1 & 0 & 1 \\
 1 & 1 & 0 \\
\end{array}
 & \iota  & 
 \\
 \cline{1-7} 7 & 2 \text{B}_4^{\text{}}\oplus\text{D}_8^{\text{}} & \mathbb{Z}_2^2  & 
\begin{array}{@{}c@{}c@{}c@{}}
 0 & 1 & c \\
 1 & 0 & c \\
\end{array}
 & \mathbb{Z}_2^2  & 
\begin{array}{@{}c@{}c@{}c@{}}
 0 & 0 & c \\
 1 & 1 & s \\
\end{array}
 & \iota  & 
 \\
\hline	\end{tabular}	
        \end{table}\begin{table}[H]\centering
		\def\arraystretch{1.1}
		\begin{tabular}{|>{$}c<{$}||>{$}c<{$}|>{$}c<{$}>{\scriptsize$}c<{$}|>{$}c<{$}>{\scriptsize$}c<{$}|>{$}c<{$}|>{$}c<{$}|}\hline A & (O_A)_{\text{root}} & \multicolumn{2}{|c|}{$\frac{O_A}{(O_A)_{\text{root}}}$}& \multicolumn{2}{|c|}{\text{Glue code}}& I & N_{I,\mathcal{G}} \\ \hline
 \hline 18 & 12 \text{A}_1^{\text{}} & \mathbb{Z}_2  & 
\begin{array}{@{}c@{}c@{}c@{}c@{}c@{}c@{}c@{}c@{}c@{}c@{}c@{}c@{}}
 1 & 1 & 1 & 1 & 1 & 1 & 1 & 1 & 1 & 1 & 1 & 1 \\
\end{array}
 & \mathbb{Z}_2^{11}  & 
\begin{array}{@{}c@{}c@{}c@{}c@{}c@{}c@{}c@{}c@{}c@{}c@{}c@{}c@{}}
 0 & 0 & 0 & 0 & 0 & 0 & 0 & 0 & 0 & 0 & 1 & 1 \\
 0 & 0 & 0 & 0 & 0 & 0 & 0 & 0 & 0 & 1 & 0 & 1 \\
 0 & 0 & 0 & 0 & 0 & 0 & 0 & 0 & 1 & 0 & 0 & 1 \\
 0 & 0 & 0 & 0 & 0 & 0 & 0 & 1 & 0 & 0 & 0 & 1 \\
 0 & 0 & 0 & 0 & 0 & 0 & 1 & 0 & 0 & 0 & 0 & 1 \\
 0 & 0 & 0 & 0 & 0 & 1 & 0 & 0 & 0 & 0 & 0 & 1 \\
 0 & 0 & 0 & 0 & 1 & 0 & 0 & 0 & 0 & 0 & 0 & 1 \\
 0 & 0 & 0 & 1 & 0 & 0 & 0 & 0 & 0 & 0 & 0 & 1 \\
 0 & 0 & 1 & 0 & 0 & 0 & 0 & 0 & 0 & 0 & 0 & 1 \\
 0 & 1 & 0 & 0 & 0 & 0 & 0 & 0 & 0 & 0 & 0 & 1 \\
 1 & 0 & 0 & 0 & 0 & 0 & 0 & 0 & 0 & 0 & 0 & 1 \\
\end{array}
 & \psi  & $\multirow{12}{*}{$[12   \text{A}_1 \, |\mathbb{Z}_2]$}$ \\
 \cline{1-7} 19 & 6 \text{C}_2^{\text{}} & \mathbb{Z}_2  & 
\begin{array}{@{}c@{}c@{}c@{}c@{}c@{}c@{}}
 1 & 1 & 1 & 1 & 1 & 1 \\
\end{array}
 & \mathbb{Z}_2^5  & 
\begin{array}{@{}c@{}c@{}c@{}c@{}c@{}c@{}}
 0 & 0 & 0 & 0 & 1 & 1 \\
 0 & 0 & 0 & 1 & 0 & 1 \\
 0 & 0 & 1 & 0 & 0 & 1 \\
 0 & 1 & 0 & 0 & 0 & 1 \\
 1 & 0 & 0 & 0 & 0 & 1 \\
\end{array}
 & \tau  &  \\
 \cline{1-7} 20 & 4 \text{B}_3^{\text{}} & \mathbb{Z}_2  & 
\begin{array}{@{}c@{}c@{}c@{}c@{}}
 1 & 1 & 1 & 1 \\
\end{array}
 & \mathbb{Z}_2^3  & 
\begin{array}{@{}c@{}c@{}c@{}c@{}}
 0 & 0 & 1 & 1 \\
 0 & 1 & 0 & 1 \\
 1 & 0 & 0 & 1 \\
\end{array}
 & \xi  &  \\
\hline	\end{tabular}	
        \end{table}\begin{table}[H]\centering
		\def\arraystretch{1.1}
		\begin{tabular}{|>{$}c<{$}||>{$}c<{$}|>{$}c<{$}>{\scriptsize$}c<{$}|>{$}c<{$}>{\scriptsize$}c<{$}|>{$}c<{$}|>{$}c<{$}|}\hline A & (O_A)_{\text{root}} & \multicolumn{2}{|c|}{$\frac{O_A}{(O_A)_{\text{root}}}$}& \multicolumn{2}{|c|}{\text{Glue code}}& I & N_{I,\mathcal{G}} \\ \hline
 \hline 35 & 3 \text{A}_2^{\text{(3)}} & \mathbb{Z}_3^2  & 
\begin{array}{@{}c@{}c@{}c@{}}
 0 & 1 & 2 \\
 1 & 0 & 2 \\
\end{array}
 & \mathbb{Z}_3  & 
\begin{array}{@{}c@{}c@{}c@{}}
 1 & 1 & 1 \\
\end{array}
 & o & [9  \text{A}_2 \, | \mathbb{Z}_3^2] \\
 \hline 36 & 3 \text{A}_2^{\text{}} & 1 & 
\begin{array}{@{}c}
 \text{} \\
\end{array}
 & \mathbb{Z}_3^3  & 
{{}}
 & \chi  & [9  \text{A}_2 \, | \mathbb{Z}_3^3] \\
 \hline 37 & \text{A}_3^{\text{(4)}}\oplus\text{C}_7^{\text{}} & \mathbb{Z}_4  & 
\begin{array}{@{}c@{}c@{}}
 1 & 1 \\
\end{array}
 & \mathbb{Z}_2  & 
\begin{array}{@{}c@{}c@{}}
 2 & 1 \\
\end{array}
 & \theta  & $\multirow{7}{*}{$[2 \text{A}_1\oplus 4 \text{A}_3 \, |\mathbb{Z}_4]$}$ \\
 \cline{1-7} 38 & \text{A}_2^{\text{(4)}}\oplus\text{C}_2^{\text{(2)}}\oplus\text{E}_6^{\text{}} & \mathbb{Z}_6  & 
\begin{array}{@{}c@{}c@{}c@{}}
 1 & 1 & 1 \\
\end{array}
 & \mathbb{Z}_3  & 
\begin{array}{@{}c@{}c@{}c@{}}
 1 & 0 & 1 \\
\end{array}
 & \lambda  &    \\
 \cline{1-7} 39 & 2 \text{A}_1^{\text{(4)}}\oplus\text{C}_3^{\text{}}\oplus\text{D}_5^{\text{}} & \mathbb{Z}_2^2  & 
\begin{array}{@{}c@{}c@{}c@{}c@{}}
 0 & 1 & 1 & 2 \\
 1 & 0 & 1 & 2 \\
\end{array}
 & \mathbb{Z}_2 \mathbb{Z}_4  & 
\begin{array}{@{}c@{}c@{}c@{}c@{}}
 1 & 1 & 1 & 0 \\
 0 & 0 & 1 & 1 \\
\end{array}
 & \pi  &    \\
 \cline{1-7} 40 & \text{A}_1^{\text{(2)}}\oplus3 \text{A}_3^{\text{}} & \mathbb{Z}_2  & 
\begin{array}{@{}c@{}c@{}c@{}c@{}}
 1 & 2 & 2 & 2 \\
\end{array}
 & \mathbb{Z}_4^3  & 
\begin{array}{@{}c@{}c@{}c@{}c@{}}
 0 & 0 & 1 & 1 \\
 0 & 1 & 0 & 1 \\
 1 & 0 & 0 & 1 \\
\end{array}
 & \varphi  &    \\
\hline
	\end{tabular}
	\caption{Orbit lattices $O_A \hookrightarrow N_I$ for cyclic 
 orbifolds}\label{tab:ocyclic}
\end{table}

\begin{table}[H]
	\centering
		\def\arraystretch{1.1}
		\begin{tabular}
   {|>{$}c<{$}||>{$}c<{$}|>{$}c<{$}>{\scriptsize$}c<{$}|>{$}c<{$}>{\scriptsize$}c<{$}|>{$}c<{$}|>{$}c<{$}|}\hline A & (O_A)_{\text{root}} & \multicolumn{2}{|c|}{$\frac{O_A}{(O_A)_{\text{root}}}$}&
   \multicolumn{2}{|c|}{\text{Glue code}}& I & N_{I,\mathcal{G}}    \\ \hline
\hline 60 & 3 \text{C}_2^{\text{(2)}}\oplus\text{D}_6^{\text{}} & \mathbb{Z}_2^3  & 
\begin{array}{@{}c@{}c@{}c@{}c@{}}
 0 & 0 & 1 & 0 \\
 0 & 1 & 0 & 0 \\
 1 & 0 & 0 & 0 \\
\end{array}
 & \mathbb{Z}_2^2  & 
\begin{array}{@{}c@{}c@{}c@{}c@{}}
 0 & 0 & 0 & s \\
 0 & 0 & 0 & c \\
\end{array}
 & \xi  & $\multirow{26}{*}{$[12 \text{A}_1 \, | \mathbb{Z}_2^2]$}$ \\
 \cline{1-7} 61 & 2 \text{C}_3^{\text{(2)}}\oplus\text{C}_6^{\text{}} & \mathbb{Z}_2^2  & 
\begin{array}{@{}c@{}c@{}c@{}}
 0 & 1 & 0 \\
 1 & 0 & 0 \\
\end{array}
 & \mathbb{Z}_2  & 
\begin{array}{@{}c@{}c@{}c@{}}
 0 & 0 & 1 \\
\end{array}
 & \iota  &  \\
 \cline{1-7} 62 & 4 \text{A}_1^{\text{(4)}}\oplus2 \text{D}_4^{\text{}} & \mathbb{Z}_2^4  & 
\begin{array}{@{}c@{}c@{}c@{}c@{}c@{}c@{}}
 0 & 0 & 0 & 1 & 0 & 0 \\
 0 & 0 & 1 & 0 & 0 & 0 \\
 0 & 1 & 0 & 0 & 0 & 0 \\
 1 & 0 & 0 & 0 & 0 & 0 \\
\end{array}
 & \mathbb{Z}_2^4  & 
\begin{array}{@{}c@{}c@{}c@{}c@{}c@{}c@{}}
 0 & 0 & 0 & 0 & 0 & s \\
 0 & 0 & 0 & 0 & 0 & c \\
 0 & 0 & 0 & 0 & s & 0 \\
 0 & 0 & 0 & 0 & c & 0 \\
\end{array}
 & \tau  &  \\
 \cline{1-7} 63 & 2 \text{C}_2^{\text{(2)}}\oplus2 \text{C}_4^{\text{}} & \mathbb{Z}_2^2  & 
\begin{array}{@{}c@{}c@{}c@{}c@{}}
 0 & 1 & 0 & 0 \\
 1 & 0 & 0 & 0 \\
\end{array}
 & \mathbb{Z}_2^2  & 
\begin{array}{@{}c@{}c@{}c@{}c@{}}
 0 & 0 & 0 & 1 \\
 0 & 0 & 1 & 0 \\
\end{array}
 & \xi  &  \\
 \cline{1-7} 64 & \text{C}_4^{\text{(2)}}\oplus2 \text{F}_4^{\text{}} & \mathbb{Z}_2  & 
\begin{array}{@{}c@{}c@{}c@{}}
 1 & 0 & 0 \\
\end{array}
 & 1 & 
\begin{array}{@{}c}
 \text{} \\
\end{array}
 & \eta  &  \\
\hline	\end{tabular}	
        \end{table}\begin{table}[H]\centering
		\def\arraystretch{1.1}
		\begin{tabular}{|>{$}c<{$}||>{$}c<{$}|>{$}c<{$}>{\scriptsize$}c<{$}|>{$}c<{$}>{\scriptsize$}c<{$}|>{$}c<{$}|>{$}c<{$}|}\hline A & (O_A)_{\text{root}} & \multicolumn{2}{|c|}{$\frac{O_A}{(O_A)_{\text{root}}}$}& \multicolumn{2}{|c|}{\text{Glue code}}& I & N_{I,\mathcal{G}} \\ \hline
\hline 79 & 2 \text{C}_5^{\text{(2)}} & \mathbb{Z}_2^2  & {{}} & 1 & 
\begin{array}{@{}c}
 \text{} \\
\end{array}
 & \varepsilon  & [14   \text{A}_1  \, | \mathbb{Z}_2^2 ] \\
 \hline 80 & 7 \text{A}_1^{\text{(2)}}\oplus\text{A}_3^{\text{}} & \mathbb{Z}_2^4  & 
\begin{array}{@{}c@{}c@{}c@{}c@{}c@{}c@{}c@{}c@{}}
 0 & 0 & 0 & 1 & 1 & 1 & 1 & 0 \\
 0 & 0 & 1 & 0 & 0 & 1 & 1 & 2 \\
 0 & 1 & 0 & 0 & 1 & 0 & 1 & 2 \\
 1 & 0 & 0 & 0 & 1 & 1 & 0 & 2 \\
\end{array}
 & \mathbb{Z}_2^3 \mathbb{Z}_4  & 
\begin{array}{@{}c@{}c@{}c@{}c@{}c@{}c@{}c@{}c@{}}
 0 & 0 & 0 & 1 & 1 & 1 & 1 & 0 \\
 0 & 1 & 1 & 0 & 0 & 1 & 1 & 0 \\
 1 & 0 & 1 & 0 & 1 & 0 & 1 & 0 \\
 0 & 0 & 1 & 0 & 1 & 1 & 0 & 1 \\
\end{array}
 & \varphi  & $\multirow{13}{*}{$[14  \text{A}_1  \, | \mathbb{Z}_2^3]$}$ \\
  \cline{1-7} 81 & 10 \text{A}_1^{\text{}} & \mathbb{Z}_2  & 
\begin{array}{@{}c@{}c@{}c@{}c@{}c@{}c@{}c@{}c@{}c@{}c@{}}
 0 & 0 & 1 & 1 & 1 & 1 & 1 & 1 & 1 & 1 \\
\end{array}
 & \mathbb{Z}_2^9  & 
\begin{array}{@{}c@{}c@{}c@{}c@{}c@{}c@{}c@{}c@{}c@{}c@{}}
 0 & 0 & 0 & 0 & 0 & 0 & 0 & 0 & 0 & 1 \\
 0 & 0 & 0 & 0 & 0 & 0 & 0 & 1 & 1 & 0 \\
 0 & 0 & 0 & 0 & 0 & 0 & 1 & 0 & 1 & 0 \\
 0 & 0 & 0 & 0 & 0 & 1 & 0 & 0 & 0 & 0 \\
 0 & 0 & 0 & 0 & 1 & 0 & 0 & 0 & 1 & 0 \\
 0 & 0 & 0 & 1 & 0 & 0 & 0 & 0 & 1 & 0 \\
 0 & 0 & 1 & 0 & 0 & 0 & 0 & 0 & 1 & 0 \\
 0 & 1 & 0 & 0 & 0 & 0 & 0 & 0 & 1 & 0 \\
 1 & 0 & 0 & 0 & 0 & 0 & 0 & 0 & 1 & 0 \\
\end{array}
 & \psi  & \\
\hline	\end{tabular}	
        \end{table}\begin{table}[H]\centering
		\def\arraystretch{1.1}
		\begin{tabular}{|>{$}c<{$}||>{$}c<{$}|>{$}c<{$}>{\scriptsize$}c<{$}|>{$}c<{$}>{\scriptsize$}c<{$}|>{$}c<{$}|>{$}c<{$}|}\hline A & (O_A)_{\text{root}} & \multicolumn{2}{|c|}{$\frac{O_A}{(O_A)_{\text{root}}}$}& \multicolumn{2}{|c|}{\text{Glue code}}& I & N_{I,\mathcal{G}} \\ \hline
 \hline 95 & 7 \text{A}_1^{\text{}} & 1 & 
\begin{array}{@{}c}
 \text{} \\
\end{array}
 & \mathbb{Z}_2^7  & {{}} & \psi  & $\multirow{4}{*}{$[17   \text{A}_1  \, |\mathbb{Z}_2^5]$}$ \\
  \cline{1-7} 96 & 5 \text{A}_1^{\text{(4)}}\oplus\text{C}_2^{\text{}} & \mathbb{Z}_2^5  & 
\begin{array}{@{}c@{}c@{}c@{}c@{}c@{}c@{}}
 0 & 0 & 0 & 0 & 1 & 0 \\
 0 & 0 & 0 & 1 & 0 & 0 \\
 0 & 0 & 1 & 0 & 0 & 0 \\
 0 & 1 & 0 & 0 & 0 & 0 \\
 1 & 0 & 0 & 0 & 0 & 0 \\
\end{array}
 & \mathbb{Z}_2  & 
\begin{array}{@{}c@{}c@{}c@{}c@{}c@{}c@{}}
 0 & 0 & 0 & 0 & 0 & 1 \\
\end{array}
 & \tau  &  \\
 \hline 97 & 6 \text{A}_1^{\text{}} & 1 & 
\begin{array}{@{}c}
 \text{} \\
\end{array}
 & \mathbb{Z}_2^6  & {{}} & \psi  & $\multirow{1}{*}{$[18   \text{A}_1  \, | \mathbb{Z}_2^6]$}$ \\
 \hline 98 & 6 \text{A}_1^{\text{(4)}} & \mathbb{Z}_2^6  & {{}} & 1 & 
\begin{array}{@{}c}
 \text{} \\
\end{array}
 & \tau  & $\multirow{2}{*}{$[18   \text{A}'_1  \, | \mathbb{Z}_2^6]$}$ \\
  \cline{1-7} 99 & 6 \text{A}_1^{\text{}} & 1 & 
\begin{array}{@{}c}
 \text{} \\
\end{array}
 & \mathbb{Z}_2^6  & {{}} & \psi  &  \\
 \hline 100 & 5 \text{A}_1^{\text{}} & 1 & 
\begin{array}{@{}c}
 \text{} \\
\end{array}
 & \mathbb{Z}_2^5  & {{}} & \psi  & [19   \text{A}_1  \, | \mathbb{Z}_2^7] \\
 \hline 101 & 4 \text{A}_1^{\text{}} & 1 & 
\begin{array}{@{}c}
 \text{} \\
\end{array}
 & \mathbb{Z}_2^4  & {{}} & \psi  & [20  \text{A}_1  \, | \mathbb{Z}_2^8] \\
\hline
   \end{tabular}
	\caption{Orbit lattices $O_A \hookrightarrow N_I$ for non-cyclic orbifolds (of rank $\geq 4$).}\label{tab:ononcyclic}
\end{table}

\begin{table}[H]
	\centering
		\def\arraystretch{1.3}
		\begin{tabular}
   {|>{$}c<{$}||>{$}c<{$}|>{$}c<{$}|>{$}c<{$}|>{$}c<{$}|>{$}c<{$}|>{$}c<{$}|}\hline \# & \Gamma_c & \text{r} &
   \mathcal{G} & \Gamma_c^\perp & \text{Gen.}
   & \text{Theory}  \\ \hline
\hline  1 & \Gamma_{\text{3,3}}\oplus\Gamma_{\text{1,1}}^{\text{(2)}}\oplus2 \text{D}_4 & 16 & \mathbb{Z}_2 & [ 8   \text{A}_1  \, | \mathbb{Z}_2] & \text{B} & \text{M on }\frac{\text{K3} \times S^1}{\mathbb{Z}_2} \\
\hline  2 & \Gamma_{\text{4,4}}^{\text{(2)}}\oplus\text{D}_4 & 12 & \mathbb{Z}_2 & [ 12   \text{A}_1  \, | \mathbb{Z}_2] & \text{D} & \text{?} \\
\hline  3 & \Gamma_{\text{4,4}}^{\text{(2)}} & 8 & \mathbb{Z}_2 & [ 16   \text{A}'_1  \, | \mathbb{Z}_2^4] & \text{} & \text{F on }S^1\times\frac{(T^4 \times S^1)'}{\mathbb{Z}_2} \\
\hline  4 & \Gamma_{\text{1,1}}\oplus\Gamma_{\text{3,3}}^{\text{(2)}} & 8 & \mathbb{Z}_2 & [ 16   \text{A}_1  \, | \mathbb{Z}_2^5] & \text{} & \text{M on }\frac{T^4 \times S^1}{\mathbb{Z}_2}\\
\hline  5 & \Gamma_{\text{3,3}}\oplus\Gamma_{\text{1,1}}^{\text{(3)}}\oplus2 \text{A}_2 & 12 & \mathbb{Z}_3 & [ 6   \text{A}_2  \, | \mathbb{Z}_3] & \text{C} & \text{M on }\frac{\text{K3} \times S^1}{\mathbb{Z}_3} \\
\hline  6 & \Gamma_{\text{2,2}}^{\text{(3)}}\oplus\text{A}_2^{\text{(-1)}} & 6 & \mathbb{Z}_3 & [ 9   \text{A}_2  \, | \mathbb{Z}_3^2] & \text{} & \text{F on }S^1\times\frac{(T^4 \times S^1)'}{\mathbb{Z}_3} \\
\hline  7 & \Gamma_{\text{1,1}}\oplus\Gamma_{\text{1,1}}^{\text{(3)}}\oplus\text{A}_2^{\text{(-1)}} & 6 & \mathbb{Z}_3 & [ 9   \text{A}_2  \, | \mathbb{Z}_3^3] & \text{} & \text{M on }\frac{T^4 \times S^1}{\mathbb{Z}_3} \\
\hline
   \end{tabular}
	\caption{Charge lattices $\Gamma_c$ for cyclic 
 orbifolds and, for the 15 known cases, examples of theories where they are realized.
 r denotes the rank and Gen. the genus on \cite{2017arXiv170805990H}.
 The primes in the F-Theory backgrounds denote a non-trivial theta angle \cite{MMHP}.
 }\label{tab:chargecyclic}
\end{table}

\begin{table}[H]
	\centering
		\def\arraystretch{1.3}
		\begin{tabular}
   {|>{$}c<{$}||>{$}c<{$}|>{$}c<{$}|>{$}c<{$}|>{$}c<{$}|>{$}c<{$}|}\hline \# & \Gamma_c & \text{r} &
   \mathcal{G} & \Gamma_c^\perp  & \text{Theory}  \\ \hline
 \hline  24 & \Gamma_{\text{2,2}}\oplus\Gamma_{\text{2,2}}^{\text{(2)}}\oplus4  \text{A}_1& 12 & \mathbb{Z}_2^2 & [12  \text{A}_1 \, |\mathbb{Z}_2^2 ] & \mathbb{Z}_2 \times \mathbb{Z}_2\text{-quadruple} \\
 \hline  25 & \Gamma_{\text{4,4}}^{\text{(2)}}\oplus2  \text{A}_1 & 10 & \mathbb{Z}_2^2 & [14  \text{A}_1 \, |\mathbb{Z}_2^2] & \text{?} \\
 \hline  26 & \Gamma_{\text{1,1}}\oplus\Gamma_{\text{3,3}}^{\text{(2)}}\oplus2  \text{A}_1 & 10 & \mathbb{Z}_2^3 & [14  \text{A}_1 \, |\mathbb{Z}_2^3] & \text{?} \\
 \hline  27 & \Gamma_{\text{4,4}}^{\text{(2)}}\oplus\text{A}_1 & 9 & \mathbb{Z}_2^3 & [15  \text{A}_1 \, |\mathbb{Z}_2^3] & \text{?} \\
 \hline  28 & \Gamma_{\text{1,1}}\oplus\Gamma_{\text{3,3}}^{\text{(2)}}\oplus\text{A}_1 & 9 & \mathbb{Z}_2^4 & [15  \text{A}_1 \, |\mathbb{Z}_2^4] & \text{?} \\
 \hline  29 & \Gamma_{\text{3,3}}^{\text{(2)}} & 8 & \mathbb{Z}_2^2 & [16  \text{A}_1 \, |\mathbb{Z}_2^4] & \text{?} \\
 \hline  30 & \Gamma_{\text{3,3}}^{\text{(2)}}\oplus\text{A}_1^{\text{(-1)}} & 7 & \mathbb{Z}_2^3 & [17  \text{A}_1 \, |\mathbb{Z}_2^5] & \text{?} \\
\hline
   \end{tabular}
	\caption{Charge lattices $\Gamma_c$ for non-cyclic orbifolds}\label{tab:chargenoncyclic}
\end{table}

\begin{table}[H]
    \vspace{-0.3in}
	\centering
		\def\arraystretch{0.95}
		\begin{minipage}[t]{0.49\textwidth}\centering
		\strut\vspace*{-\baselineskip}\newline
		\begin{tabular}
   {|>{$}c<{$}|>{\scriptsize$}c<{$}|}\hline  \text{Coinvariant lattice} & 
   \normalsize\text{Glue code}    \\ \hline
\hline [ 8   \text{A}_1  \, | \mathbb{Z}_2  ] & 
\begin{array}{@{}c@{}c@{}c@{}c@{}c@{}c@{}c@{}c@{}}
 1 & 1 & 1 & 1 & 1 & 1 & 1 & 1 \\
\end{array}\\
 \hline [ 12   \text{A}_1  \, | \mathbb{Z}_2  ] & 
\begin{array}{@{}c@{}c@{}c@{}c@{}c@{}c@{}c@{}c@{}c@{}c@{}c@{}c@{}}
 1 & 1 & 1 & 1 & 1 & 1 & 1 & 1 & 1 & 1 & 1 & 1 \\
\end{array}\\
 \hline [ 12   \text{A}_1  \, | \mathbb{Z}_2^2  ] & 
\begin{array}{@{}c@{}c@{}c@{}c@{}c@{}c@{}c@{}c@{}c@{}c@{}c@{}c@{}}
 0 & 0 & 0 & 0 & 1 & 1 & 1 & 1 & 1 & 1 & 1 & 1 \\
 1 & 1 & 1 & 1 & 0 & 0 & 0 & 0 & 1 & 1 & 1 & 1 \\
\end{array}\\
 \hline [ 14   \text{A}_1  \, | \mathbb{Z}_2^2  ] & 
\begin{array}{@{}c@{}c@{}c@{}c@{}c@{}c@{}c@{}c@{}c@{}c@{}c@{}c@{}c@{}c@{}}
 0 & 0 & 0 & 0 & 0 & 0 & 1 & 1 & 1 & 1 & 1 & 1 & 1 & 1 \\
 1 & 1 & 1 & 1 & 1 & 1 & 0 & 0 & 0 & 0 & 0 & 0 & 1 & 1 \\
\end{array}\\
 \hline [ 14   \text{A}_1  \, | \mathbb{Z}_2^3  ] & 
\begin{array}{@{}c@{}c@{}c@{}c@{}c@{}c@{}c@{}c@{}c@{}c@{}c@{}c@{}c@{}c@{}}
 0 & 0 & 0 & 0 & 0 & 0 & 1 & 1 & 1 & 1 & 1 & 1 & 1 & 1 \\
 0 & 0 & 1 & 1 & 1 & 1 & 0 & 0 & 0 & 0 & 1 & 1 & 1 & 1 \\
 1 & 1 & 0 & 0 & 1 & 1 & 0 & 0 & 1 & 1 & 0 & 0 & 1 & 1 \\
\end{array}\\
 \hline [ 15   \text{A}_1  \, | \mathbb{Z}_2^3  ] & 
\begin{array}{@{}c@{}c@{}c@{}c@{}c@{}c@{}c@{}c@{}c@{}c@{}c@{}c@{}c@{}c@{}c@{}}
 0 & 0 & 0 & 0 & 0 & 0 & 0 & 1 & 1 & 1 & 1 & 1 & 1 & 1 & 1 \\
 0 & 0 & 0 & 1 & 1 & 1 & 1 & 0 & 0 & 0 & 0 & 1 & 1 & 1 & 1 \\
 1 & 1 & 1 & 0 & 1 & 1 & 1 & 0 & 0 & 0 & 1 & 0 & 0 & 0 & 1 \\
\end{array}\\
 \hline [ 15   \text{A}_1  \, | \mathbb{Z}_2^4  ] & 
\begin{array}{@{}c@{}c@{}c@{}c@{}c@{}c@{}c@{}c@{}c@{}c@{}c@{}c@{}c@{}c@{}c@{}}
 0 & 0 & 0 & 0 & 0 & 0 & 0 & 1 & 1 & 1 & 1 & 1 & 1 & 1 & 1 \\
 0 & 0 & 0 & 1 & 1 & 1 & 1 & 0 & 0 & 0 & 0 & 1 & 1 & 1 & 1 \\
 0 & 1 & 1 & 0 & 0 & 1 & 1 & 0 & 0 & 1 & 1 & 0 & 0 & 1 & 1 \\
 1 & 0 & 1 & 0 & 1 & 0 & 1 & 0 & 1 & 0 & 1 & 0 & 1 & 0 & 1 \\
\end{array}\\
\hline\end{tabular}
		\end{minipage}
		\begin{minipage}[t]{0.50\textwidth}\centering
		\strut\vspace*{-\baselineskip}\newline
  \begin{tabular}
		  {|>{$}c<{$}|>{\scriptsize$}c<{$}|}\hline  \text{Coinvariant lattice} & \normalsize\text{Glue code}    \\ \hline
 \hline [ 6   \text{A}_2  \, | \mathbb{Z}_3  ] & 
\begin{array}{@{}c@{}c@{}c@{}c@{}c@{}c@{}}
 1 & 1 & 2 & 2 & 2 & 2 \\
\end{array}\\
 \hline [ 8   \text{A}_2  \, | \mathbb{Z}_3^2  ] & 
\begin{array}{@{}c@{}c@{}c@{}c@{}c@{}c@{}c@{}c@{}}
 0 & 0 & 1 & 1 & 1 & 1 & 1 & 1 \\
 1 & 2 & 0 & 0 & 1 & 1 & 2 & 2 \\
\end{array}\\
 \hline [ 9   \text{A}_2  \, | \mathbb{Z}_3^2  ] & 
\begin{array}{@{}c@{}c@{}c@{}c@{}c@{}c@{}c@{}c@{}c@{}}
 0 & 0 & 0 & 1 & 1 & 1 & 1 & 2 & 2 \\
 1 & 2 & 2 & 0 & 0 & 2 & 2 & 0 & 1 \\
\end{array}\\
 \hline [ 9   \text{A}_2  \, | \mathbb{Z}_3^3  ] & 
\begin{array}{@{}c@{}c@{}c@{}c@{}c@{}c@{}c@{}c@{}c@{}}
 0 & 0 & 0 & 1 & 1 & 1 & 2 & 2 & 2 \\
 0 & 1 & 1 & 0 & 1 & 2 & 0 & 1 & 2 \\
 1 & 0 & 2 & 0 & 1 & 1 & 1 & 1 & 0 \\
\end{array}\\
 \hline [ 10   \text{A}_2  \, | \mathbb{Z}_3^4  ] & 
\begin{array}{@{}c@{}c@{}c@{}c@{}c@{}c@{}c@{}c@{}c@{}c@{}}
 0 & 0 & 0 & 0 & 1 & 1 & 1 & 1 & 1 & 2 \\
 0 & 0 & 1 & 2 & 0 & 0 & 1 & 2 & 2 & 2 \\
 0 & 1 & 0 & 2 & 0 & 2 & 2 & 0 & 1 & 2 \\
 1 & 0 & 0 & 2 & 0 & 2 & 1 & 1 & 0 & 1 \\
\end{array}\\
 \hline [  2   \text{A}_1 \oplus 4   \text{A}_3 \, |  \mathbb{Z}_4  ] & 
\begin{array}{@{}c@{}c@{}c@{}c@{}c@{}c@{}}
 1 & 1 & 1 & 1 & 1 & 3 \\
\end{array}\\
 \hline [  4   \text{A}_1 \oplus 4   \text{A}_3 \, |  \mathbb{Z}_2 \mathbb{Z}_4  ] & 
\begin{array}{@{}c@{}c@{}c@{}c@{}c@{}c@{}c@{}c@{}}
 1 & 1 & 1 & 1 & 0 & 0 & 2 & 2 \\
 0 & 0 & 1 & 1 & 1 & 1 & 1 & 3 \\
\end{array}\\
\hline\end{tabular}
		\end{minipage}   \caption{Coinvariant lattices $N_{I,\mathcal{G}}$ specified by their glue code, which is equivalent to $N_{I,\mathcal{G}}/(N_{I,\mathcal{G}})_\text{root}$ in virtue of always having roots of norm 2.}\label{tab:sing}
\end{table}

\begin{table}[H]
	\centering
		\def\arraystretch{1.3}
		\begin{tabular}
   {|c||>{$}c<{$}||>{$}c<{$}|>{$}c<{$}|>{$}c<{$}|>{$}c<{$}|>{$}c<{$}|}\hline d & \# & \Gamma_c & \text{r} &
   \mathcal{G} & \Gamma_c^\perp  & \text{Theory}  \\ \hline
\hline  \multirow{12}{*}{7}& 1 & \Gamma_{\text{3,3}}\oplus 2 \text{D}_4 & 14 & \mathbb{Z}_2 & [ 2   \text{D}_4  ] &  \text{F on }\frac{\text{K3} \times S^1}{\mathbb{Z}_2} \\
\cline{2-7} &  3 & \Gamma_{\text{3,3}}^{\text{(2)}} & 6 & \mathbb{Z}_2 & [ 4   \text{D}_4  \, | \mathbb{Z}_2] &\text{F on } \frac{(T^4 \times S^1)'}{\mathbb{Z}_2} \\
\cline{2-7} &  4 & \Gamma_{\text{1,1}}\oplus\Gamma_{\text{2,2}}^{\text{(2)}} & 6 & \mathbb{Z}_2 & [ 4   \text{D}_4  \, | \mathbb{Z}_2^2] & \text{F on }\frac{T^4 \times S^1}{\mathbb{Z}_2}\\
\cline{2-7} &  5 & \Gamma_{\text{3,3}}\oplus 2 \text{A}_2 & 10 & \mathbb{Z}_3 & [ 2   \text{E}_6 ] & \text{F on }\frac{\text{K3} \times S^1}{\mathbb{Z}_3} \\
\cline{2-7} &  6 & \Gamma_{\text{1,1}}^{\text{(3)}}\oplus\text{A}_2^{\text{(-1)}} & 4 & \mathbb{Z}_3 & [ 3   \text{E}_6 ] & \text{F on } \frac{(T^4 \times S^1)'}{\mathbb{Z}_3} \\
\cline{2-7} &  7 & \Gamma_{\text{1,1}}\oplus\text{A}_2^{\text{(-1)}} & 4 & \mathbb{Z}_3 & [ 3   \text{E}_6  \, | \mathbb{Z}_3] & \text{F on }\frac{T^4 \times S^1}{\mathbb{Z}_3} \\
\cline{2-7} &  8 & \Gamma_{\text{3,3}}\oplus 2 \text{A}_1 & 8 & \mathbb{Z}_4 & [2   \text{E}_7] &  \text{F on }\frac{\text{K3} \times S^1}{\mathbb{Z}_4} \\
\hline   \end{tabular}
	\caption{Charge lattices $\Gamma_c$ for 7, 8 and 9 dimensions with examples of theories where they are realized. \# denotes the number in Table \ref{tab:chargecyclic} of the dimensional reduced theory in 6d. The asterisks mean circle compactifications of the corresponding 9D theories at the bottom of the table. Both 9D theories with $\# = 4$ compactify become dual when compactified to 8D.   
 }\label{tab:otherdim}
\end{table}

\newpage
\bibliographystyle{JHEP}
\bibliography{Niemeier6d}

\providecommand{\href}[2]{#2}\begingroup\raggedright\begin{thebibliography}{10}

\bibitem{Kim:2019ths}
H.-C. Kim, H.-C. Tarazi, and C.~Vafa, {\it {Four-dimensional
  $\mathbf{\mathcal{N}=4}$ SYM theory and the swampland}},  {\em Phys. Rev. D}
  {\bf 102} (2020), no.~2 026003, [\href{http://arxiv.org/abs/1912.06144}{{\tt
  arXiv:1912.06144}}].

\bibitem{Montero:2020icj}
M.~Montero and C.~Vafa, {\it {Cobordism Conjecture, Anomalies, and the String
  Lamppost Principle}},  {\em JHEP} {\bf 01} (2021) 063,
  [\href{http://arxiv.org/abs/2008.11729}{{\tt arXiv:2008.11729}}].

\bibitem{Hamada:2021bbz}
Y.~Hamada and C.~Vafa, {\it {8d Supergravity, Reconstruction of Internal
  Geometry and the Swampland}},  \href{http://arxiv.org/abs/2104.05724}{{\tt
  arXiv:2104.05724}}.

\bibitem{Bedroya:2021fbu}
A.~Bedroya, Y.~Hamada, M.~Montero, and C.~Vafa, {\it {Compactness of Brane
  Moduli and the String Lamppost Principle in $d>6$}},
  \href{http://arxiv.org/abs/2110.10157}{{\tt arXiv:2110.10157}}.

\bibitem{Cvetic:2020kuw}
M.~Cveti\v{c}, M.~Dierigl, L.~Lin, and H.~Y. Zhang, {\it {String Universality
  and Non-Simply-Connected Gauge Groups in 8d}},  {\em Phys. Rev. Lett.} {\bf
  125} (2020), no.~21 211602, [\href{http://arxiv.org/abs/2008.10605}{{\tt
  arXiv:2008.10605}}].

\bibitem{Lee:2021usk}
S.-J. Lee, W.~Lerche, and T.~Weigand, {\it {Physics of Infinite Complex
  Structure Limits in eight Dimensions}},
  \href{http://arxiv.org/abs/2112.08385}{{\tt arXiv:2112.08385}}.

\bibitem{Lee:2021qkx}
S.-J. Lee and T.~Weigand, {\it {Elliptic K3 Surfaces at Infinite Complex
  Structure and their Refined Kulikov models}},
  \href{http://arxiv.org/abs/2112.07682}{{\tt arXiv:2112.07682}}.

\bibitem{Collazuol:2022jiy}
V.~Collazuol, M.~Gra\~na, and A.~Herr\'aez, {\it {$E_9$ symmetry in the
  Heterotic String on $S^1$ and the Weak Gravity Conjecture}},
  \href{http://arxiv.org/abs/2203.01341}{{\tt arXiv:2203.01341}}.

\bibitem{Vafa:2005ui}
C.~Vafa, {\it {The String landscape and the swampland}},
  \href{http://arxiv.org/abs/hep-th/0509212}{{\tt hep-th/0509212}}.

\bibitem{Ooguri:2006in}
H.~Ooguri and C.~Vafa, {\it {On the Geometry of the String Landscape and the
  Swampland}},  {\em Nucl. Phys. B} {\bf 766} (2007) 21--33,
  [\href{http://arxiv.org/abs/hep-th/0605264}{{\tt hep-th/0605264}}].

\bibitem{Kachru:2016ttg}
S.~Kachru, N.~M. Paquette, and R.~Volpato, {\it {3D String Theory and Umbral
  Moonshine}},  {\em J. Phys. A} {\bf 50} (2017), no.~40 404003,
  [\href{http://arxiv.org/abs/1603.07330}{{\tt arXiv:1603.07330}}].

\bibitem{Gaberdiel:2011fg}
M.~R. Gaberdiel, S.~Hohenegger, and R.~Volpato, {\it {Symmetries of K3 sigma
  models}},  {\em Commun. Num. Theor. Phys.} {\bf 6} (2012) 1--50,
  [\href{http://arxiv.org/abs/1106.4315}{{\tt arXiv:1106.4315}}].

\bibitem{Cheng:2016org}
M.~C.~N. Cheng, S.~M. Harrison, R.~Volpato, and M.~Zimet, {\it {K3 String
  Theory, Lattices and Moonshine}},
  \href{http://arxiv.org/abs/1612.04404}{{\tt arXiv:1612.04404}}.

\bibitem{Harrison:2020wxl}
S.~M. Harrison, N.~M. Paquette, D.~Persson, and R.~Volpato, {\it {Fun with
  $F_{24}$}},  {\em JHEP} {\bf 02} (2021) 039,
  [\href{http://arxiv.org/abs/2009.14710}{{\tt arXiv:2009.14710}}].

\bibitem{Harrison:2021gnp}
S.~M. Harrison, N.~M. Paquette, D.~Persson, and R.~Volpato, {\it {BPS Algebras
  in 2D String Theory}},  \href{http://arxiv.org/abs/2107.03507}{{\tt
  arXiv:2107.03507}}.

\bibitem{Persson:2015jka}
D.~Persson and R.~Volpato, {\it {Fricke S-duality in CHL models}},  {\em JHEP}
  {\bf 12} (2015) 156, [\href{http://arxiv.org/abs/1504.07260}{{\tt
  arXiv:1504.07260}}].

\bibitem{Persson:2017lkn}
D.~Persson and R.~Volpato, {\it {Dualities in CHL-Models}},  {\em J. Phys. A}
  {\bf 51} (2018), no.~16 164002, [\href{http://arxiv.org/abs/1704.00501}{{\tt
  arXiv:1704.00501}}].

\bibitem{MMHP}
M.~Montero and H.~P. de~Freitas, {\it {New Supersymmetric String Theories from
  Discrete Theta Angles}},  \href{http://arxiv.org/abs/2209.03361}{{\tt
  arXiv:2209.03361}}.

\bibitem{deBoer:2001wca}
J.~de~Boer, R.~Dijkgraaf, K.~Hori, A.~Keurentjes, J.~Morgan, D.~R. Morrison,
  and S.~Sethi, {\it {Triples, fluxes, and strings}},  {\em Adv. Theor. Math.
  Phys.} {\bf 4} (2002) 995--1186,
  [\href{http://arxiv.org/abs/hep-th/0103170}{{\tt hep-th/0103170}}].

\bibitem{HP}
H.~P. De~Freitas, {\it {New Supersymmetric String Moduli Spaces from Frozen
  Singularities}},  \href{http://arxiv.org/abs/2209.03451}{{\tt
  arXiv:2209.03451}}.

\bibitem{Fraiman:2018ebo}
B.~Fraiman, M.~Gra\~na, and C.~A. N\'u\~nez, {\it {A new twist on heterotic
  string compactifications}},  {\em JHEP} {\bf 09} (2018) 078,
  [\href{http://arxiv.org/abs/1805.11128}{{\tt arXiv:1805.11128}}].

\bibitem{Font:2020rsk}
A.~Font, B.~Fraiman, M.~Gra\~na, C.~A. N\'u\~nez, and H.~P. De~Freitas, {\it
  {Exploring the landscape of heterotic strings on $T^d$}},  {\em JHEP} {\bf
  10} (2020) 194, [\href{http://arxiv.org/abs/2007.10358}{{\tt
  arXiv:2007.10358}}].

\bibitem{Font:2021uyw}
A.~Font, B.~Fraiman, M.~Gra\~na, C.~A. N\'u\~nez, and H.~Parra De~Freitas, {\it
  {Exploring the landscape of CHL strings on $T^d$}},
  \href{http://arxiv.org/abs/2104.07131}{{\tt arXiv:2104.07131}}.

\bibitem{Fraiman:2021hma}
B.~Fraiman and H.~P. de~Freitas, {\it {Freezing of gauge symmetries in the
  heterotic string on T$^{4}$}},  {\em JHEP} {\bf 04} (2022) 007,
  [\href{http://arxiv.org/abs/2111.09966}{{\tt arXiv:2111.09966}}].

\bibitem{Fraiman:2021soq}
B.~Fraiman and H.~P. De~Freitas, {\it {Symmetry enhancements in 7d heterotic
  strings}},  {\em JHEP} {\bf 10} (2021) 002,
  [\href{http://arxiv.org/abs/2106.08189}{{\tt arXiv:2106.08189}}].

\bibitem{Cvetic:2021sjm}
M.~{Cvetic}, M.~{Dierigl}, L.~{Lin}, and H.~Y. {Zhang}, {\it {On the Gauge
  Group Topology of 8d CHL Vacua}},  {\em arXiv e-prints} (July, 2021)
  arXiv:2107.04031, [\href{http://arxiv.org/abs/2107.04031}{{\tt
  arXiv:2107.04031}}].

\bibitem{Cvetic:2022uuu}
M.~Cvetic, M.~Dierigl, L.~Lin, and H.~Y. Zhang, {\it {One Loop to Rule Them
  All: Eight and Nine Dimensional String Vacua from Junctions}},
  \href{http://arxiv.org/abs/2203.03644}{{\tt arXiv:2203.03644}}.

\bibitem{Schweigert:1996tg}
C.~Schweigert, {\it {On moduli spaces of flat connections with nonsimply
  connected structure group}},  {\em Nucl. Phys. B} {\bf 492} (1997) 743--755,
  [\href{http://arxiv.org/abs/hep-th/9611092}{{\tt hep-th/9611092}}].

\bibitem{Lerche:1997rr}
W.~Lerche, C.~Schweigert, R.~Minasian, and S.~Theisen, {\it {A Note on the
  geometry of CHL heterotic strings}},  {\em Phys. Lett. B} {\bf 424} (1998)
  53--59, [\href{http://arxiv.org/abs/hep-th/9711104}{{\tt hep-th/9711104}}].

\bibitem{Fuchs:1995zr}
J.~Fuchs, B.~Schellekens, and C.~Schweigert, {\it {From Dynkin diagram
  symmetries to fixed point structures}},  {\em Commun. Math. Phys.} {\bf 180}
  (1996) 39--98, [\href{http://arxiv.org/abs/hep-th/9506135}{{\tt
  hep-th/9506135}}].

\bibitem{Schellekens:1992db}
A.~N. Schellekens, {\it {Meromorphic C = 24 conformal field theories}},  {\em
  Commun. Math. Phys.} {\bf 153} (1993) 159--186,
  [\href{http://arxiv.org/abs/hep-th/9205072}{{\tt hep-th/9205072}}].

\bibitem{2017arXiv170805990H}
G.~{H{\"o}hn}, {\it {On the Genus of the Moonshine Module}},  {\em arXiv
  e-prints} (Aug., 2017) arXiv:1708.05990,
  [\href{http://arxiv.org/abs/1708.05990}{{\tt arXiv:1708.05990}}].

\bibitem{fp2022}
B.~Fraiman and H.~Parra De~Freitas, ``Symmetry enhancements in string vacua
  with 16 supercharges.'' \url{https://bernardofraiman.github.io/16SUSY/},
  2022.

\bibitem{Narain:1986qm}
K.~S. Narain, M.~H. Sarmadi, and C.~Vafa, {\it {Asymmetric Orbifolds}},  {\em
  Nucl. Phys. B} {\bf 288} (1987) 551.

\bibitem{Ginsparg:1986bx}
P.~H. Ginsparg, {\it {Comment on Toroidal Compactification of Heterotic
  Superstrings}},  {\em Phys. Rev. D} {\bf 35} (1987) 648.

\bibitem{Polchinski:1995df}
J.~Polchinski and E.~Witten, {\it {Evidence for heterotic - type I string
  duality}},  {\em Nucl. Phys. B} {\bf 460} (1996) 525--540,
  [\href{http://arxiv.org/abs/hep-th/9510169}{{\tt hep-th/9510169}}].

\bibitem{Chaudhuri:1995fk}
S.~Chaudhuri, G.~Hockney, and J.~D. Lykken, {\it {Maximally supersymmetric
  string theories in D \ensuremath{<} 10}},  {\em Phys. Rev. Lett.} {\bf 75}
  (1995) 2264--2267, [\href{http://arxiv.org/abs/hep-th/9505054}{{\tt
  hep-th/9505054}}].

\bibitem{Chaudhuri:1995bf}
S.~Chaudhuri and J.~Polchinski, {\it {Moduli space of CHL strings}},  {\em
  Phys. Rev. D} {\bf 52} (1995) 7168--7173,
  [\href{http://arxiv.org/abs/hep-th/9506048}{{\tt hep-th/9506048}}].

\bibitem{Mikhailov:1998si}
A.~Mikhailov, {\it {Momentum lattice for CHL string}},  {\em Nucl. Phys. B}
  {\bf 534} (1998) 612--652, [\href{http://arxiv.org/abs/hep-th/9806030}{{\tt
  hep-th/9806030}}].

\bibitem{zbMATH00993515}
K.-i. Nishiyama, {\it The {Jacobian} fibrations on some {{\(K3\)}} surfaces and
  their {Mordell}-{Weil} groups},  {\em Jpn. J. Math., New Ser.} {\bf 22}
  (1996), no.~2 293--347.

\bibitem{zbMATH01236805}
K.-i. Nishiyama, {\it A remark on {Jacobian} fibrations on {{\(K\)}}3
  surfaces},  {\em Saitama Math. J.} {\bf 15} (1997) 67--71.

\bibitem{Dabholkar:1998kv}
A.~Dabholkar and J.~A. Harvey, {\it {String islands}},  {\em JHEP} {\bf 02}
  (1999) 006, [\href{http://arxiv.org/abs/hep-th/9809122}{{\tt
  hep-th/9809122}}].

\bibitem{GS}
A.~{Garbagnati} and A.~{Sarti}, {\it {Elliptic fibrations and symplectic
  automorphisms on K3 surfaces}},  {\em arXiv e-prints} (Jan., 2008)
  arXiv:0801.3992, [\href{http://arxiv.org/abs/0801.3992}{{\tt
  arXiv:0801.3992}}].

\bibitem{Acharya:2022shu}
B.~S. Acharya, G.~Aldazabal, A.~Font, K.~Narain, and I.~G. Zadeh, {\it
  {Heterotic Strings on ${\mathbb T^3}/{\mathbb Z_2}$, Nikulin involutions and
  M-theory}},  \href{http://arxiv.org/abs/2205.09764}{{\tt arXiv:2205.09764}}.

\bibitem{Cheng:2022nso}
P.~Cheng, I.~V. Melnikov, and R.~Minasian, {\it {Flat equivariant gerbes:
  holonomies and dualities}},  \href{http://arxiv.org/abs/2207.06885}{{\tt
  arXiv:2207.06885}}.

\bibitem{Fischer:2012qj}
M.~Fischer, M.~Ratz, J.~Torrado, and P.~K.~S. Vaudrevange, {\it {Classification
  of symmetric toroidal orbifolds}},  {\em JHEP} {\bf 01} (2013) 084,
  [\href{http://arxiv.org/abs/1209.3906}{{\tt arXiv:1209.3906}}].

\bibitem{Fischer:2013qza}
M.~Fischer, S.~Ramos-Sanchez, and P.~K.~S. Vaudrevange, {\it {Heterotic
  non-Abelian orbifolds}},  {\em JHEP} {\bf 07} (2013) 080,
  [\href{http://arxiv.org/abs/1304.7742}{{\tt arXiv:1304.7742}}].

\end{thebibliography}\endgroup

\end{document}